\documentclass[11pt]{article}
\usepackage{amsfonts,slashed}
\usepackage{arydshln}
\usepackage[greek,english]{babel}
\usepackage{cancel}
\usepackage{upgreek}
\usepackage{float}
\usepackage{url}
\usepackage{latexsym}
\usepackage{amsfonts}
\usepackage{amsmath}
\usepackage{epsfig}
\usepackage{latexsym,amssymb}
\usepackage{amsmath,amssymb,amsthm,mathtools}
\usepackage{hyperref}
\usepackage{fontenc}
\usepackage{graphicx,tikz}
\usepackage{multirow}
\usepackage{xfrac}
\usepackage[T1]{fontenc}
\usepackage{textcomp}
\usepackage{lmodern}
\usepackage{stmaryrd}
\usepackage{cite}
\usepackage[normalem]{ulem}

\usepackage[vcentermath]{youngtab}

\newdimen\squaresize \squaresize=12pt
\newdimen\thickness \thickness=0.7pt

\def\square#1{\hbox{\vrule width \thickness
   \vbox to \squaresize{\hrule height \thickness\vss
      \hbox to \squaresize{\hss#1\hss}
   \vss\hrule height\thickness}
\unskip\vrule width \thickness} \kern-\thickness}

\def\cut#1{\hbox{\vrule width-1 \thickness
   \vbox to \squaresize{\hrule height-1 \thickness\vss
      \hbox to \squaresize{\hss#1\hss}
   \vss\hrule height-1\thickness}
\unskip\vrule width +4 \thickness} \kern-\thickness}

\def\vsquare#1{\vbox{\square{$#1$}}\kern-\thickness}

\def\young#1{
\vbox{\smallskip\offinterlineskip \halign{&\vsquare{##}\cr #1}}}

\newcommand{\tinyyoung}[1]{
\squaresize=7pt \thickness=0.4pt \mbox{\scriptsize\young{#1}}
\squaresize=15pt \thickness=0.7pt}

\usepackage{stmaryrd}

\usepackage[margin=20pt,small]{caption}
\usepackage{subcaption}

\usepackage{ifpdf}
\ifx\pdfoutput\undefined
   \pdffalse
   \usepackage{cite}
 \else
   \pdfoutput=1
   \pdftrue
\fi
\DeclareFontFamily{OMS}{rsfs}{\skewchar\font'60}
\DeclareFontShape{OMS}{rsfs}{m}{n}{<-5>rsfs5 <5-7>rsfs7 <7->rsfs10 }{}
\DeclareSymbolFont{rsfs}{OMS}{rsfs}{m}{n}
\DeclareSymbolFontAlphabet{\Scr}{rsfs}

\numberwithin{equation}{section}
\def\be{\begin{equation}}
\def\ee{\end{equation}}
\def\ba{\begin{array}}
\def\ea{\end{array}}

\newcommand{\bea}{\begin{eqnarray}}
\newcommand{\eea}{\end{eqnarray}}

\newcommand{\ft}[2]{{\textstyle\frac{#1}{#2}}}

\def\={~=~}
\def\*{{}^*}

\usepackage[shadow,textwidth=2.7cm]{todonotes}
\usepackage{ifthen}
\reversemarginpar
\newcounter{todocounter}

\colorlet{fccolor}{blue!40!white}

\newcommand{\fcinline}[2][]{
  \ifthenelse { \equal {#1} {} }
    { \def\temp {#2} }  % if #1 == blank
    { \def\temp {#1} }   % else (not blank)
  \refstepcounter{todocounter}\todo[color=fccolor,inline,caption={\textbf{\thetodocounter. FC} \temp}]{\textbf{\thetodocounter. FC:} #2}{}}

\colorlet{hscolor}{orange!20!white}

\newcommand{\hsinline}[2][]{
  \ifthenelse { \equal {#1} {} }
    { \def\temp {#2} }  % if #1 == blank
    { \def\temp {#1} }   % else (not blank)
  \refstepcounter{todocounter}\todo[color=hscolor,inline,caption={\textbf{\thetodocounter. HS} \temp}]{\textbf{\thetodocounter. HS:} #2}{}}

\def\={~=~}
\def\*{{}^*}

\begin{document}
\begin{titlepage}
\vfill
\begin{flushright}
\end{flushright}

\bigskip

\begin{center}
	{\LARGE \bf A supergravity dual for IKKT holography}\\[1cm]
	
	{\large\bf Franz Ciceri\,$^{a}{\!}$
		\footnote{\tt franz.ciceri@ens-lyon.fr} and Henning Samtleben\,${}^{a,b}{\!}$
		\footnote{\tt henning.samtleben@ens-lyon.fr} \vskip .8cm}	
		{\it ${}^a$ ENSL, CNRS, Laboratoire de physique, F-69342 Lyon, France}\\[2ex] \ 
	{\it  $^{b}$ Institut Universitaire de France (IUF)}\\ 
	
\end{center}
\vspace{0.5cm}
%%%%%%%%%%%%%%%%%%%%%%%%%%%%%%%%%%%%%%%%%%%%%%%

\noindent

\begin{abstract}
\noindent 
The IKKT matrix model, from the holographic perspective,  arises at the $p=-1$ endpoint of the family of dualities relating type II supergravities on near-horizon D$p$-brane geometries to $(p+1)$-dimensional super Yang-Mills theories with sixteen supercharges. In this work, we detail and expand results reported in a recent letter by establishing the holographic dictionary between gauge-invariant operators in the lowest BPS multiplet of the matrix model and the corresponding Kaluza-Klein fluctuations of Euclidean IIB supergravity compactified on the D$(-1)$ instanton background. The full non-linear dynamics of these fluctuations can be encoded in a one-dimensional maximal gauged supergravity, which we construct explicitly. We provide its complete Lagrangian, up to second order in fermions, together with the corresponding supersymmetry transformations. We further discuss real forms of the theory for non-compact gauge groups and their embeddings into ten-dimensional supergravities with Lorentzian signature. From the analysis of the one-dimensional Killing spinor equations, we derive different classes of half-supersymmetric solutions, and discuss their uplifts as well as their relations to known solutions.

\end{abstract}

\setcounter{footnote}{0}

\end{titlepage}

%%%%%%%%%%%%%%%%%%%%%%%%%%%%%%%%%%%%%%%%%%%%%%%

\tableofcontents \noindent {}

%%%%%%%%%%%%%%%%%%%%%%%%%%%%%%%%%%%%%%%%%%%%%%%

\section {Introduction}\label{sec:intro}

Holographic dualities continue to serve as one of the most powerful tools for probing the non-perturbative regime of gauge theories. Among the wide landscape of such dualities, a particularly rich although technically challenging sector arises from the correspondence between the near-horizon backgrounds of D$p$-branes and the non-conformal field theories living on their worldvolumes. In contrast to the maximally symmetric AdS/CFT correspondence, where the conformal symmetry severely constrains the dynamics on both sides of the duality, the D$p$-brane systems exhibit scale-dependent behavior closer to realistic quantum field theories. Although non-conformal, these dualities retain sixteen supercharges. On the field-theory side, they are governed by the 
reduction of ten-dimensional $\mathrm{SU}(N)$ super Yang–Mills to $p+1$ dimensions, which is conformal only in the special case 
$p=3$. 

This article provides a detailed account of the results reported in \cite{Ciceri:2025maa}, pushing this correspondence to the extremal limit $p=-1$, where
the worldvolume theory reduces to the zero-dimensional IKKT matrix model which has by itself
been proposed as a candidate for a nonperturbative
formulation of type IIB superstring theory \cite{Ishibashi:1996xs}. 
The dual supergravity background in this case is the half-supersymmetric D$(-1)$ instanton solution of Euclidean IIB supergravity with a flat spacetime metric but non-trivial dilaton/axion \cite{Gibbons:1995vg,Gubser:1996wt,Bergshoeff:1998ry,Ooguri:1998pf}.
The study of non-conformal dualities goes back to the early days of holography \cite{Itzhaki:1998dd,Boonstra:1998mp} and has evolved into a well-defined computational framework \cite{Bianchi:2001kw, Wiseman:2008qa,Kanitscheider:2008kd}, enabling the quantitative analysis of such dualities. Recent tests of these dualities for correlation functions for general $p$ have been reported in \cite{Bobev:2025idz}.

Over the years, the IKKT matrix model has been extensively studied using a variety of numerical and analytical methods, providing important insights into its structure and dynamics. Early Monte Carlo simulations highlighted the crucial role of supersymmetry in ensuring convergence of the partition function and suggested a striking dependence on the arithmetic divisor function $\sigma_2(N)$ \cite{Krauth:1998xh}, as initially conjectured in \cite{Green:1997tn} and later confirmed for arbitrary $N$ via an exact computation using supersymmetric localization \cite{Moore:1998et}. Subsequent numerical studies \cite{Ambjorn:2000dx,Nishimura:2011xy,Anagnostopoulos:2022dak} also provided evidence for the spontaneous breaking of the
$\mathrm{SO}(10)$ R-symmetry and for the dynamical emergence of spacetime structures. Despite these compelling results, the IKKT model has remained, until recently, largely unexplored through the lens of holography. Yet this perspective offers a new angle on the large $N$ limit of the model and on the long-standing question of how space and time can emerge from pure matrix integrals. In the past few years, however, the maximally supersymmetric mass deformation of the IKKT model \cite{Bonelli:2002mb} has attracted considerable attention, enabling the first concrete tests of so-called `timeless holography’ \cite{Hartnoll:2024csr,Komatsu:2024bop,Komatsu:2024ydh,Hartnoll:2025ecj,Chou:2025rwy}. 

In this paper, we detail and expand results reported in \cite{Ciceri:2025maa} and establish the holographic dictionary between single trace operators in the lowest BPS multiplet of the matrix model and the corresponding Kaluza-Klein fluctuations of Euclidean IIB supergravity compactified on the D$(-1)$ instanton background. The full non-linear dynamics of these fluctuations can be encoded in a one-dimensional maximal gauged supergravity, which we present in detail.

Concretely, the rest of this paper is organized as follows. In Section~\ref{sec:IKKT}, we outline the IKKT model and the spectrum of its ${\rm SU}(N)$ invariant single trace operators. We review the results of \cite{Morales:2004xc}, showing that this spectrum organizes into infinite towers of long and short supermultiplets. The latter are the BPS multiplets which are in one-to-one correspondence with the fluctuations of the dual IIB supergravity around the round sphere $S^9$. We focus on the lowest BPS multiplet which combines 129 bosonic and 128 fermionic degrees of freedom, after properly accounting for the `gauge' degrees of freedom of ${\rm SO}(10)$ R-symmetry and supersymmetry. We spell out the corresponding operators and their supersymmetry transformation rules. The main part of the paper is then devoted to the construction of the bulk realization of this multiplet. In analogy to the higher-dimensional cases \cite{Boonstra:1998mp} this leads to a maximal ${\rm SO}(10)$-gauged supergravity theory in one dimension. 

In Section~\ref{sec:CT}, we start by constructing a bosonic subsector of this theory, which can be directly obtained by consistent truncation of the ten-dimensional dilaton-axion system on a nine-sphere $S^9$, upon properly adapting the standard techniques of \cite{deWit:1986oxb,Nastase:2000tu,Cvetic:2000dm,Ciceri:2023bul}. The scalar target space of the resulting one-dimensional theory is given by a gauged coset space ${\rm SL}(10)/{\rm SO}(10)_K$, coupled to one-dimensional dilaton-gravity. In one dimension, neither gravity nor vector fields have kinetic terms, but couple algebraically through an einbein and gauge fields acting as Lagrange multipliers. We give the explicit uplift formulas to ten, and, following \cite{Tseytlin:1996ne}, further to twelve dimensions. Finally, we discuss in detail the generalization of the construction to internal hyperboloid spaces and different spacetime signatures. The former implies a change of the gauge group ${\rm SO}(10)_\mathrm{g} \rightarrow {\rm SO}(p,q)_\mathrm{g}$, while the latter induces a different (indefinite) scalar target space ${\rm SL}(10)/{\rm SO}(s,t)_K$, of the one-dimensional theory.

In Section~\ref{sec:SUGRA}, we construct the full maximally supersymmetric one-dimensional theory, by adding the full field content of the supermultiplet, and imposing supersymmetry of the action. In particular, the bosonic field content of the previous subsector is enhanced by 120 axion fields, which are expected to arise from reduction of the ten-dimensional p-form fields. We determine the action and supersymmetry transformation rules to quadratic order in the fermions and in particular determine the scalar potential which is a fourth order polynomial in the axion fields. While most of the discussion is focused on the ${\rm SO}(10)_\mathrm{g}$-gauged model, descending from reduction of the (complex) Euclidean IIB supergravity, we also discuss in some detail the real forms of the theory which feature a scalar target space ${\rm SL}(10)/{\rm SO}(9,1)_K$ and are expected to uplift to the exotic IIB' and IIB$^*$ theories of \cite{Hull:1998vg,Hull:1998ym}.

Finally, in section~\ref{sec:BPS}, we study 1/2-BPS solutions of the one-dimensional theory. In absence of axion fields, we determine the most general 1/2-BPS solution which is characerized by 10 real constants. We present the uplift to a ten-dimensional solution with flat space-time metric. For coinciding constants, this solution reduces to the half-supersymmetric D$(-1)$ instanton solution of Euclidean IIB supergravity \cite{Gibbons:1995vg,Gubser:1996wt,Bergshoeff:1998ry,Ooguri:1998pf}. With non-vanishing axion fields we reproduce the 1/2-BPS spherical brane solutions of \cite{Bobev:2018ugk,Bobev:2024gqg}, preserving an ${\rm SO}(7)$ subgroup of the gauge group.

We conclude with three appendices, that collect conventions and some technical details.

%%%%%%%%%%%%%%%%%%%%%%%%%%%%%%%%%%%%%%%%%%%%%%%

\section{IKKT model}\label{sec:IKKT}

In this section, we review the definition and some properties of the IKKT matrix model, proposed in \cite{Ishibashi:1996xs} as a non-perturbative definition of IIB superstring theory. 
The model is given by an action
\begin{equation}
S_{\text{\tiny{IKKT}}}=
-
\text{Tr}\Big[\tfrac14[X_a,X_b][X^a,X^b]-\tfrac12\bar{\Psi}\,\Gamma^a\,[X_a,\Psi]\Big]\,,
\label{eq:IKKTaction}
\end{equation}
where, $X_a$ and $\Psi^{\alpha}$ are bosonic and fermionic ${\mathfrak su}(N)$-valued matrices, respectively.
Vector indices $a=1, \dots, 10$, are contracted in Euclidean signature, and the $\Gamma_a$ denote the $\mathrm{SO}(10)$ $\Gamma$-matrices. The spinors $\Psi^{\alpha}$, $\alpha=1, \dots, 32$, are complex and satisfy the chirality condition
\bea
\Gamma_{*}\Psi=\Psi\;,\qquad
\Gamma_{*}\equiv-i\,\Gamma_{1}\Gamma_2\ldots\Gamma_{10}
\,.
\label{eq:chirality}
\eea  
We summarize our spinor conditions in Appendix~(\ref{app:gamma}), in particular we choose a basis in which the chirality matrix $\Gamma_{*}$ is block diagonal.
In (\ref{eq:IKKTaction}), we suppress explicit spinor indices, and define the charge-conjugated spinors as 
\begin{equation}
    \bar \Psi_\alpha:=\Psi^\beta\mathcal C_{\alpha\beta}
    \,,\label{eq:CCIKKT}
\end{equation}
with the charge conjugation matrix 
$\mathcal C=-i\Gamma_{10}\Gamma_{*}$, c.f.\ (\ref{eq:CC}). In particular, the 
Hermitian conjugate of $\Psi^\alpha$ does not appear in the (complex) action $S_{\text{\tiny{IKKT}}}$. This is a standard property of spinorial actions in Euclidean signature \cite{Nicolai:1978vc,Krauth:1998xh}.
The above formulae match those of \cite{Komatsu:2024ydh} and \cite{Hartnoll:2024csr} up to a change of conventions for the gamma and charge conjugation matrices. Let us note that the Lorentzian version of the IKKT sigma model is obtained by sending
\begin{equation}
    \Gamma_{10} \rightarrow -i\,\Gamma_{10}
    \,,\;\;\;\;\;{X_{10}\rightarrow -i\,X_{10}\,,}
\end{equation}
replacing ${\rm SO}(10)$ by ${\rm SO}(9,1)$. Accordingly, the chirality matrix is given by $\Gamma_{*}\equiv\Gamma_{1}\Gamma_2\ldots\Gamma_{10}$ and spinors can be equipped with a reality condition, such that the action (\ref{eq:IKKTaction}) is real. The Lorentzian model is obtained by dimensional reduction of ten-dimensional super Yangs-Mills theory down to zero dimensions.

The equations of motion that follow from varying the action (\ref{eq:IKKTaction}) with respect to $X_a$ and $\Psi$ are algebraic, and take the form
\bea
\big[X_b,[X^a,X^b]\big]+\frac12\{\bar\Psi,\Gamma^a\Psi\}=0\,,
\nonumber\\
\Gamma^a\big[\Psi,X_a\big]=0\,.
\label{eq:eomIKKT}
\eea
The action (\ref{eq:IKKTaction}) is invariant under ${\rm SO}(10)$, ${\rm SU}(N)$, and supersymmetry transformations.\footnote{Although all these transformations are global symmetries in zero dimensions, we still refer to ${\rm SU}(N)$ as `gauge symmetries' in accordance with their higher-dimensional origin.} The latter act as
\bea
\delta_\epsilon X^a&=&\bar\epsilon\, \Gamma^a\,\Psi\,,
\nonumber\\
\delta_\epsilon \Psi&=&\frac12\,\Gamma^{ab}\,\epsilon\,[X_a,X_b]\,,
\label{eq:susy}
\eea
where the  SUSY parameter $\epsilon$ satisfies the same chirality condition (\ref{eq:chirality}). 
The supersymmetry algebra closes on-shell into
\begin{equation}
[\delta_{\epsilon_1},\delta_{\epsilon_2}]
=\delta^{{\rm SU}(N)}_{\lambda}\,,\label{eq:IKKETalgebra}
\end{equation}
with the ${\rm SU}(N)$ parameter $\lambda=2\,\bar\epsilon_2\,\Gamma^a\,\epsilon_1X_a$\,.
Closure on the fermions $\Psi$ requires Fierzing and the fermionic field equations from (\ref{eq:eomIKKT}).
We also note the scaling symmetry
\begin{equation}
    X^a \rightarrow \lambda \,X^a\,,\qquad\Psi^\alpha\rightarrow \lambda^{3/2}\,\Psi^\alpha\,,
    \label{eq:scaling}
\end{equation}
under which the action (\ref{eq:IKKTaction}) scales homogeneously.

%%%%%%%%%%%%%%%%%%%%%%%%%%%%%%%%%%%%%%%%%%%%%%%
\subsection{Single trace operators and Polya counting}\label{sec:Polya}
%%%%%%%%%%%%%%%%%%%%%%%%%%%%%%%%%%%%%%%%%%%%%%%

In this section, we will discuss the `gauge invariant', i.e.\ ${\rm SU}(N)$ singlet operators of the model. As in the higher dimensional gauge/gravity dualities, we expect the supergravity states to be dual to certain single trace operators, built as
\begin{equation}
{\cal O} = {\rm Tr}\left[ XX\dots\Psi\dots X \dots \Psi \dots \right]
\,.
\label{eq:single-trace}
\end{equation}
from the basic fields $X^a$ and $\Psi^\alpha$. The latter transform under ${\rm SO}(10)$ as 
\begin{equation}
{\bf 10}_1 \ominus {\bf 16}_{s}{}_{\frac32}
\,.
\label{eq:XPsi}
\end{equation}
The subscript here refers to the charge under rescaling (\ref{eq:scaling}), and $\ominus$ refers to the direct sum, however keeping track of the fermionic statistics of the spinor components.
The representation content and structure of the gauge invariant single-trace operators (\ref{eq:single-trace}) has been computed in \cite{Morales:2004xc}.
Enumerating the set of independent single trace operators amounts to counting the cyclic words in an alphabet made from the letters $\{X^a, \Psi^\alpha\}$ and subtracting the algebraic field equations (\ref{eq:eomIKKT}). Let us illustrate this at the lowest levels.

Single letter operators ${\rm Tr}[X^a]$,  ${\rm Tr}[\Psi^\alpha]$ vanish since we are considering ${\rm SU}(N)$ matrices. The lowest gauge invariant operators are thus built from two letters and fill the symmetric tensor product
\begin{equation}
 \tinyyoung{&\cr} 
 \,:\;\;    \big({\bf 10}_{1} \ominus {\bf 16}_{s}{}_{\frac32}\big) \otimes_{\rm sym} \big({\bf 10}_{1} \ominus {\bf 16}_{s}{}_{\frac32}\big)
    \,.
    \label{eq:l2}
\end{equation}
which expands into
\begin{equation}
 \tinyyoung{&\cr} 
 \,:\;\;{\bf 1}_2 \oplus {\bf 54}_2 \ominus
    {\bf 16}_{c}{}_{\frac52} \ominus {\bf 144}_{c}{}_{\frac52} \oplus {\bf 120}_3 
    \,.
    \label{eq:prod2}
\end{equation}
These representations capture all operators built from two letter: ${\rm Tr}[X^aX^b]$, ${\rm Tr}[X^a\Psi^\alpha]$, and ${\rm Tr}[\Psi^\alpha \Psi^\beta]$. 

At three letters, the gauge invariant operators fill the tensor product
\begin{equation}
  \tinyyoung{&&\cr} \;\, \oplus  \tinyyoung{\cr\cr\cr} 
 \,:\;\;
    \big({\bf 10}_{1} \ominus {\bf 16}_{s}{}_{\frac32}\big)^{{\otimes 3}_{\rm sym}}
    \;\oplus\; \big({\bf 10}_{1} \ominus {\bf 16}_{s}{}_{\frac32}\big)^{{\otimes 3}_{\rm antisym}}
    \,,
\end{equation}
since three-letter words with hook symmetry $ \tinyyoung{&\cr\cr}$
are incompatible with the cyclicity of the trace. The full representation content of these tensor products is given by
\bea
  \tinyyoung{&&\cr} 
 &:&
     {\bf 10}_3
   \oplus    {\bf 210}_3
  \ominus  {\bf 16}_{s}{}_{\frac72}
    \ominus   {\bf 144}_{c}{}_{\frac72}
   \ominus  {\bf 720}_{s}{}_{\frac72}
     \oplus  {\bf 210}_4
     \oplus  {\bf 45}_4
     \oplus  {\bf 945}_4
    \ominus  {\bf 560}_{c}{}_{\frac92}
    \,,
\nonumber\\
  \tinyyoung{\cr\cr\cr} 
   &:&
     {\bf 120}_3
    \ominus  {\bf 16}_{s}{}_{\frac72}
     \ominus  {\bf 560}_{s}{}_{\frac72}
     \ominus {\bf 144}_{c}{}_{\frac72}
    \oplus     {\bf 1}_4
    \oplus   {\bf 210}_4
    \oplus   {\bf 45}_4
     \oplus  {\bf 1050}_{s\,4}
    \oplus   {\bf 54}_4
    \nonumber\\
    &&{}
  \ominus    {\bf 144}_{s}{}_{\frac92}
   \ominus   {\bf 672}_{s}{}_{\frac92}
   \,.
    \label{eq:prod3}
\eea
In order to identify the non-trivial operators, one further needs to subtract all operators that are constructed by contracting the field equations (\ref{eq:eomIKKT}) transforming in 
\begin{equation}
 {\bf 10}_3   \ominus    {\bf 16}_{c}{}_{5/2}  
 \,,
 \label{eq:Reom}
\end{equation}
with a single letter (\ref{eq:XPsi}) before tracing over ${\rm SU}(N)$ indices.
This yields the representations
\bea
 \ominus  {\bf 16}_{c}{}_{\frac52}
\otimes
\big({\bf 10}_{1} \ominus {\bf 16}_{s}{}_{\frac32}\big)
&=&
     \ominus {\bf 16}_{s}{}_{\frac72}
     \ominus {\bf 144}_{c}{}_{\frac72}
    \oplus {\bf 1}_4
     \oplus {\bf 45}_4
     \oplus {\bf 210}_4
     \,,
\nonumber\\
{\bf 10}_3  
\otimes
\big({\bf 10}_{1} \ominus {\bf 16}_{s}{}_{\frac32}\big)
&=&
      {\bf 1}_4
     \oplus {\bf 45}_4
     \oplus {\bf 54}_4
     \ominus {\bf 16}_{c}{}_{\frac92}
     \ominus {\bf 144}_{s}{}_{\frac92}
\,,
\label{eq:over1}
\eea
to be subtracted.
The two ${\bf 45}_{4}$ on the r.h.s.\ of the two equations correspond to two different cubic structures of type ${\rm Tr}[X\Psi\Psi]$ obtained by projection of the field equations which must both be subtracted. In contrast, it is important to note that among the cubic terms there is only one singlet structure ${\rm Tr}[X^a\bar\Psi\Gamma_a\Psi]$ which can 
be obtained either by contraction of the bosonic field equations in (\ref{eq:eomIKKT}) with a X, or alternatively by contraction of the fermionic field equations with an $\Psi$. Its appearance in both lines of (\ref{eq:over1}) thus is an overcounting of subtractions that must be corrected.\footnote{The appearance of the ${\bf 16}_{c}{}_{\frac92}$ in (\ref{eq:over1}) may appear surprising as no such contribution arises in the three-letter words (\ref{eq:prod2}). Closer inspection shows that the corresponding cubic term in the contraction of the bosonic field equations in (\ref{eq:eomIKKT}) with a $\Psi$ vanishes, such that this contribution only becomes visible in the quartic terms.}

Putting everything together, a concise counting of single trace gauge invariant operators over all lengths is achieved by using Polya theory \cite{PolyaRead,Polyakov:2001af,Bianchi:2003wx} to count the cyclic words constructed from an alphabet given by the partition function
\begin{equation}
{\bf Z_1}(t)\equiv {\bf 10}\, t -{\bf 16}_s\, t^{3/2} +{\bf 16}_c\,
t^{5/2}-{\bf 10}\, t^3+{\bf 1}\, t^4
\;,
\label{eq:z1} 
\end{equation}
where we keep track of the scaling charge by the weight variable $t$.
In this single letter partition function, the first two representations correspond to the elementary single letter fields (\ref{eq:XPsi}), the two subsequent representations implement the subtraction of the field equations (\ref{eq:Reom}), while the final ${\bf 1}_4$ corrects for the overcounting of subtractions as discussed after (\ref{eq:over1}).
Cyclic words in the alphabet (\ref{eq:z1}) are counted by Polya's formula
\begin{equation}
{\cal Z}_{\rm IKKT}(t)
=
  -\sum_{m=1}^\infty \frac{\varphi(m)}{m}\,\log\left[1-{\bf Z_1}(t^m) \right]
  \;,
\label{eq:ZIKKT}
\end{equation}
with Euler's totient function $\varphi(m)$. The full ${\rm SO}(10)$ representation content can be extracted from this formula upon replacing the ${\bf 10}$, ${\bf 16}$, in (\ref{eq:z1}) by the corresponding character polynomials, c.f.\ formula (\ref{eq:ZIKKT-CP}).

%%%%%%%%%%%%%%%%%%%%%%%%%%%%%%%%%%%%%%%%%%%%%%%
\subsection{BPS multiplets}
%%%%%%%%%%%%%%%%%%%%%%%%%%%%%%%%%%%%%%%%%%%%%%%
The sum (\ref{eq:ZIKKT}) has been evaluated in \cite{Morales:2004xc} where it was also shown that the resulting operators fit into infinite towers of long and short super-multiplets. 
The supersymmetry algebra (\ref{eq:IKKETalgebra}) shows that supersymmetry transformations anti-commute on gauge invariant operators (\ref{eq:single-trace}).
Long multiplets are obtained in the standard way by the unconstrained action of all supercharges, transforming in the ${\bf 16}_c$, onto a given ${\rm SO}(10)$ representation ${\cal R}$,
with the full representation content
\begin{equation}
{\cal R} \otimes 
\left(
{\bf 1}\ominus {\bf 16}_c \oplus\, {\bf 16}_c \wedge {\bf 16}_c \,\ominus\, {\bf 16}_c \wedge {\bf 16}_c \wedge {\bf 16}_c \,\oplus\, \dots
\right)
\;,
\label{eq:long}
\end{equation}
carrying $2^{15}\cdot {\rm dim}\,{\cal R}$ bosonic and $2^{15}\cdot {\rm dim}\,{\cal R}$ fermionic states, respectively.

Apart from the long multiplets (\ref{eq:long}), the sum (\ref{eq:ZIKKT}) also contains an infinite tower of short BPS multiplets
\begin{equation}
\bigoplus_{n=2}^\infty {\cal B}_n
\,.
\label{eq:towerBPS}
\end{equation}
Their respective lowest weight state are particular operators (\ref{eq:single-trace}) of the type
\begin{equation}
    {\cal O}^{a_1 \dots a_n} =
    {\rm Tr}\big[ X^{(\!(a_1}X^{a_2} \dots X^{a_n)\!)} \big]
    \,,
\end{equation}
where $(\!( \dots)\!)$ denotes traceless symmetrization, such that the operator transforms in the $[n,0000]$ representation of ${\rm SO}(10)$, see Appendix~\ref{app:SO10} for our ${\rm SO}(10)$ notations.
The BPS multiplet ${\cal B}_n$
combines the ${\rm SO}(10)$ representations collected in Table~\ref{tab:BPS}. Counting their dimensions via (\ref{eq:SO10dims}), shows that the multiplet ${\cal B}_n$ carries 
\bea
n_{\rm f}=128\,{\rm dim}\,{\cal R}_{[n-2,0,0,0]}&& \mbox{fermionic, and}
\nonumber\\
n_{\rm b} = 1+128\,{\rm dim}\,{\cal R}_{[n-2,0,0,0]}
&& \mbox{bosonic},
\label{eq:dof}
\eea
degrees of freedom, where
\begin{equation}
{\rm dim}\,{\cal R}_{[n-2,0,0,0]}=\frac{1}{4\cdot 7!}\,(n-1)\,n\,(n+1)(n+2)^2(n+3)(n+4)(n+5)\,.
\end{equation}
The universal mismatch (\ref{eq:dof}) between bosonic and fermionic degrees of freedom appears somewhat mysterious but persists for all the BPS multiplets.
It may be taken as an indication that the notion of degrees of freedom becomes more subtle in low dimensions.

\begin{table}
\centering
\begin{tabular}{c}{}
$[n,0000]_n$
\nonumber\\{}
$[n\!-\!1,0001]_{n+\frac12}$
\nonumber\\{}
$ [n\!-\!2,0100]_{n+1}$
\nonumber\\{}
 $[n\!-\!3,1010]_{n+\frac32}$
\nonumber\\{}
 $[n\!-\!3,0020]_{n+2}
\,\oplus\, [n\!-\!4,2000]_{n+2}$
\nonumber\\{}
 $[n\!-\!4,1010]_{n+\frac52}$
\nonumber\\{}
 $[n\!-\!4,0100]_{n+3}$
\nonumber\\{}
 $[n\!-\!4,0001]_{n+\frac72}$
\nonumber\\{}
 $[n\!-\!4,0000]_{n+4}$
\end{tabular}
    \caption{BPS multiplet ${\cal B}_n$. Supercharges act vertically from top to bottom.}
    \label{tab:BPS}
\end{table}

For small values of $n$, the generic structure of Table~\ref{tab:BPS} 
degenerates and formally includes representations of negative Dynkin labels, which contribute as states of negative multiplicity, c.f.~(\ref{eq:negn}).
Specifically, the lowest `singleton' multiplet ${\cal B}_1$ coincides with the single letter partition function ${\bf Z}_1$ of (\ref{eq:z1}). As discussed above, the states of negative multiplicity here account for the subtraction of field equations (\ref{eq:Reom}). Note that those negative multiplicities occur with negative and positive signs for bosons and fermions, respectively.

For the BPS multiplet ${\cal B}_2$, the structure of Table~\ref{tab:BPS} 
degenerates into
\begin{align}
{\cal B}_2 =\,& 
{\bf 54}_{2} \ominus
{\bf 144}_{s\,\frac52} \oplus
{\bf 120}_{3}
\ominus
{\bf 45}_{4} 
\oplus
{\bf 16}_{c\,\frac92}
 \,,
 \label{eq:BPS2}
\end{align}
consistent with the counting of dimensions (\ref{eq:dof}).
Here, the states of negative multiplicity account for the global ${\rm SO}(10)$ symmetry and supersymmetry, respectively. In zero dimensions, these symmetries can be used to eliminate scalar and spinor degrees of freedom, respectively.
The first three terms in (\ref{eq:BPS2}) correspond to the explicit operators
\begin{align}
{\cal O}^{ab} =\,& {\rm Tr}[X^{a} X^{b}]-\tfrac1{10}\,\delta^{ab}\,{\rm Tr}[X^{c} X^{c}] 
\,,\nonumber\\
{\cal O}^a =\,& {\rm Tr}[X^a\,\Psi] - \tfrac19\, {\rm Tr}[X_b\,\Gamma^{ab} \Psi]
\,,\nonumber\\
{\cal O}^{abc} =\,& {\rm Tr}\left[X^{a} [X^{b},X^c]\right] -\tfrac18\, {\rm Tr}\left[\bar\Psi \Gamma^{abc} \Psi \right]
\,,\label{eq:O123}
\end{align}
where the relative coefficients in the first two operators are determined by irreducibility of the ${\rm SO}(10)$ representations, while the relative coefficient in ${\cal O}^{abc}$ is fixed by supersymmetry. Specifically, 
under supersymmetry (\ref{eq:susy}) these operators transform into each other as
\begin{align}
\delta_\epsilon \mathcal O^{ab}=\,&\tfrac95\, \Gamma^{(a}\mathcal O^{b)}\,,\nonumber\\
    \delta_\epsilon \mathcal O^a=\,&\tfrac{1}{18}\left(
    7\,\Gamma_{bc}\,\epsilon \,\mathcal O^{abc}-\Gamma^{abcd}\,\epsilon\,\mathcal O_{bcd}\right)
    \,,\nonumber\\
   \delta_\epsilon {\cal O}^{abc} =\,&0
   \,,
\end{align}
consistent with the nilpotent structure of the supersymmetry algebra (\ref{eq:IKKETalgebra}) on gauge invariant operators.

In analogy to the higher-dimensional holographic dualities one expects the states in the tower of BPS multiplets 
(\ref{eq:towerBPS}) to be dual to the fluctuations of IIB supergravity around the near-horizon $S^9$ geometry.
Indeed, counting the spherical harmonics around the coset space $S^9={\rm SO}(10)/{\rm SO}(9)$ by standard Kaluza-Klein techniques \cite{Salam:1981xd}, the IIB fluctuations can be shown to precisely match the field content of  (\ref{eq:towerBPS}) \cite{Morales:2004xc}.
In particular, the lowest BPS multiplet ${\cal B}_2$ (\ref{eq:BPS2}) in the IKKT spectrum is realized as the lowest Kaluza-Klein supergravity multiplet in the bulk. In the higher-dimensional dualities, the dynamics of this lowest multiplet is described by a maximal gauged supergravity in $(p+2)$ dimensions, with gauge group ${\rm SO}(9-p)$, \cite{Boonstra:1998mp}. Here, the associated bulk theory for IKKT corresponds to a one-dimensional maximal ${\rm SO}(10)$-gauged supergravity~\cite{Ciceri:2025maa}. It combines the lowest Kaluza-Klein fluctuations around the D$(-1)$-instanton background, realizing the field content of (\ref{eq:BPS2}).
In the following sections, we give a detailed account of the construction of \cite{Ciceri:2025maa}.

%%%%%%%%%%%%%%%%%%%%%%%%%%%%%%%%%%%%%%%%%%%%%%%

\section{Consistent truncation of IIB supergravity on $S^9$}\label{sec:CT}

In this and the following section, we will construct the one-dimensional maximal supergravity theory that describes the bulk realization of the lowest BPS multiplet.
Obtaining the full theory directly by sphere compactification of IIB supergravity is still out of reach due to the high degree of non-linearity of the expected reduction formulae. A systematic derivation of such formulae would require the embedding of the reduction into some exceptional field theory, as has been realized in other dimensions~\cite{Baguet:2015sma,Bossard:2022wvi,Bossard:2023jid}. Instead, we will proceed by explicitly constructing in a first step the consistent truncation of a subsector of IIB supergravity on the sphere $S^9$ along the lines of similar constructions in other dimensions~\cite{Nastase:2000tu,Cvetic:2000dm,Ciceri:2023bul}. Within the full BPS multiplet (\ref{eq:BPS2}), this construction will account for  ${\bf 54}$ scalar fields together with ${\bf 45}$ vector fields. The latter act as constraints or Lagrange multipliers in the one-dimensional theory, in accordance with their negative multiplicity in (\ref{eq:BPS2}).
In the next section, imposing maximal supersymmetry in one dimension, we will then extend the one-dimensional theory to a maximal supergravity, realizing the full field content of the multiplet (\ref{eq:BPS2}).

\subsection{Consistent truncation}

Our starting point is the bosonic Lagrangian 
\begin{equation}
{\cal L}_{10} =
 |E|\Big(R - \tfrac12 \partial_\mu \Phi \,\partial^\mu\Phi  
+\tfrac12 e^{2\Phi}\, \partial_\mu \mathcal X \,\partial^\mu \mathcal X \Big) \;,
\label{eq:LEu}
\end{equation}
in ten-dimensional Euclidean spacetime. It describes the axion/dilaton sector of IIB supergravity.\footnote{The corresponding action in the Einstein frame is defined as $S_{10}=\frac{1}{(2\pi)^7g_s^2l_s^8}\int d^{10}x\,\mathcal L_{10}$, where $l_s$ and $g_s$ denote the string length and the string coupling constant, respectively.} Note that in Euclidean signature the sign of the axion kinetic term changes. Consequently, the coset parametrized by $\Phi$ and $\mathcal X$ is $\rm{SL}(2)/\rm{SO}(1,1)$. 
Throughout this section, we treat all ten-dimensional fields of (\ref{eq:LEu}) as real. In the full Euclidean IIB supergravity, the reality properties of fields become more subtle, as Euclidean signature requires a selfdual 5-form to be complex, and the absence of Majorana-Weyl spinors requires a complex spinorial action. We will come back to this in the next section, when we extend the construction to a full maximally supersymmetric theory. Let us note here already, that in other spacetime signatures (specifically $(9,1)$ and $ (5,5)$), the Lagrangian (\ref{eq:LEu}) can be embedded into the exotic IIB' and IIB$^*$ theories
\cite{Hull:1998vg,Hull:1998ym}, in which all fields can be taken to be real, but certain kinetic terms exhibit sign flips, see \cite{Bergshoeff:2007cg,DHoker:2025nid}, for a careful discussion.

Euclidean IIB supergravity (\ref{eq:LEu}) admits a half-supersymmetric instanton solution 
\cite{Gibbons:1995vg,Gubser:1996wt}. The Einstein frame metric, the axion $\mathcal X$, and the dilaton $\Phi$ for this solution are given by
\begin{equation}
d s^2_{10}=dt^2+t^2\,d\Omega_9^2\,,\qquad
e^{\Phi}=H(t)=-\mathcal X^{-1} \,,
\label{eq:D-1brane}
\end{equation}
while all the other p-forms vanish. Here, 
\begin{equation}
d\Omega_9 = d\mu_i d\mu_i
\,,\quad
i=1, \dots, 10,
\label{eq:dO9}
\end{equation} 
is the line element on the sphere $S^9$ of unit radius given the embedding coordinates $\mu_i\mu_i=1$.
The metric (\ref{eq:D-1brane}) is flat with isometry group $\rm{SO}(10)$.
The harmonic function $H(t)$ in (\ref{eq:D-1brane}) reads
\begin{equation}
H=\mathrm{h}+\frac{\mathrm{Q}}{t^8}\,.\label{eq:harmonicD-1}
\end{equation}
The constant Q is related to the instanton charge,\footnote{More specifically the solution (\ref{eq:D-1brane}) describes the backreaction of $N$ coincident D($-1$)-branes located at $t=0$. Quantization of the RR 9-form flux (dual to $F_{(1)}=d\mathcal X$) through $S^9$ fixes $\mathrm{Q}=48(2\pi)^3g_s^2 Nl_s^8$.}, while the constant h can be shifted by the global ${\rm SL}(2,\mathbb{R})$ symmetry of the IIB theory \cite{Bergshoeff:1998ry}. The string frame metric describes a wormhole geometry, but becomes flat in the near-horizon limit \cite{Ooguri:1998pf}. 
The relative sign among the kinetic terms in (\ref{eq:LEu}) is essential for ensuring that the energy momentum tensor of the flat space solution \eqref{eq:D-1brane} vanishes, as required by the Einstein equations.

When expanded around the sphere $S^{9}$, the ten-dimensional theory \eqref{eq:LEu} admits a consistent truncation to a finite set of Kaluza-Klein modes described by a one-dimensional theory. The truncation ansatz is inspired by the constructions around spheres in other dimensions \cite{Nastase:2000tu,Cvetic:2000dm,Ciceri:2023bul}
 and takes the form
\begin{align}
d s_{10}^2 =\,&  e^{9\phi}\,\Delta\, \mathrm{e}^2\,dt^2 
+ {\rm g}^{-2}\, e^{\phi} \, ({T}^{-1})^{ij}\, 
{ D}\mu_i\, { D}\mu_j
\,,\nonumber\\[3mm]
e^ {\Phi} =\,&  
e^{-4\phi}\,\Delta^{-1}
\,,\nonumber\\
\mathcal X=&-\frac1{2\,{\rm g}\mathrm{e}}
\left(
(T^{-1})^{ij}\, D_t T_{kj}\,\mu_{i}\mu_{k}-\dot\phi\right)\,.\label{eq:Ansatz}
\end{align}
Here, the $\mu_i$ denote the embedding coordinates of $S^9$ from (\ref{eq:dO9}). The one-dimensional fields parametrizing the ansatz (\ref{eq:Ansatz}) are an einbein ${\rm e}$, a dilaton $\phi$, together with 54 scalar fields parametrizing a symmetric positive definite ${\rm SL}(10)$ matrix $T_{ij}$, and 45 vectors $A_{ij}=-A_{ji}$. All these fields are functions of the (Euclidean) time $t$ and derivatives are denoted by a dot, $\dot\phi=\partial_t\phi$, etc.
The vector fields only appear via covariant derivatives
\begin{align}
{ D} \mu_i =\,& d\mu_i - {\rm g}\, A_{ji}\,\mu_j\,dt\,,
\nonumber\\
 { D}_t  T_{ij} =\,&
  \dot{T}_{ij} -2\, {\rm g}\,A_{k(i}  T_{j)k}\,,
\label{eq:covD}
\end{align}
with gauge coupling constant ${\rm g}$. The warp factor $\Delta$ is given by
\begin{equation}
\Delta=T_{ij}\,\mu_i\mu_j
\,.
\end{equation}

Plugging the ansatz (\ref{eq:Ansatz}) into the ten-dimensional field equations obtained from varying (\ref{eq:LEu}), these equations reduce to a set of ordinary differential equations in $t$ for the one-dimensional fields $\{{\rm e}, \phi, T_{ij}, A_{ij}\}$
\bea
20\,\dot\phi^2 -\frac{1}{2}\, 
({T}^{-1})^{ik}({T}^{-1})^{jl}\,
D_t{T}_{kl}\,  D_t{T}_{ij}
&=&-{\rm g}^2\,{\rm e}^2\,e^{8\phi}\,
\left(
2\,T_{ij}T_{ij}-( T_{ii})^2\right),\nonumber\\
\partial_t \big({\rm e}^{-1}\,\dot\phi \big)&=&-\frac15\,{\rm g}^2 \,{\rm e}\,e^{8\phi}\,\left(
2\,T_{ij}T_{ij}-( T_{ii})^2\right),\nonumber\\
 D_t\Big({\rm e}^{-1} (T^{-1})^{ki} D_t T_{jk}\Big) % +D_t\Big({\rm e}^{-1} (T^{-1})^{kj} D_t T_{ik}\Big) 
 &=&2\,{\rm g}^2\,{\rm e}\,e^{8\phi} \,\Big(2\, T_{k\small{(\!(}i} T_{j\small{)\!)}k}-T_{\small{(\!(}ij\small{)\!)}}T_{kk}\Big)\,,
 \nonumber\\
 (T^{-1})^{ki}\, D_t T_{jk} &=& (T^{-1})^{kj}\, D_t T_{ik} \,,
 \label{eq:eomEOM}
\eea
where brackets $(\!(\dots)\!)$ denote traceless symmetrization.
In turn, these equations are obtained from variation of the one-dimensional Lagrangian
\begin{align}
\mathcal L_1 =\,& 10\,\mathrm{e}^{-1}\, \dot\phi^2-\frac14\, \mathrm{e}^{-1}\, 
({T}^{-1})^{ik}({T}^{-1})^{jl}\,
D_t{T}_{kl}\,  D_t\,{T}_{ij}
-\frac12\,\mathrm{e}\, {\rm g}^2 \,e^{8\phi}\left(
2\,T_{ij}T_{ij}-( T_{ii})^2\right)\,.
\label{eq:L1Eu}
\end{align}
Consistency of the reduction ansatz (\ref{eq:Ansatz}) is shown by straightforward computation. As an intermediate result, we give the derivative of the ten-dimensional axion in the form
\begin{equation}
F_{(1)} = d\mathcal X =
-\, {\rm g} {\rm e}\,e^{8\phi}\, U\,dt -
\frac{1}{{\rm g}{\rm e}} \, (T^{-1})^{ij}\,D_t T_{jk}\,\mu_k\,D \mu_i\,,
\label{eq:KKAEu3}
\end{equation}
with
\begin{equation}
U=2\,T_{ik}T_{jk}\mu_i\mu_j- T_{ii}\Delta
\,,
\end{equation}
which is obtained from (\ref{eq:Ansatz}) upon using the one-dimensional field equations 
(\ref{eq:eomEOM}). The dual field strength then takes the form
\begin{align}
F_{(9)}= e^{2\Phi}\,\ast\!F_{(1)}=
\frac1{9!\,{\rm g}^8}\,\Delta^{-2}\,\varepsilon^{i_1\ldots i_{10}}\Big(&U\, D \mu_{i_1}\wedge\ldots\wedge D\mu_{i_9}\,\mu_{i_{10}}\nonumber\\
&-9\,T_{ji_{10}}\,D_tT_{ki_1}\,dt \wedge D\mu_{i_2}\wedge\ldots\wedge D\mu_{i_9}\,\mu_j\,\mu_k\Big)\,.
\label{eq:dualF}
\end{align}
It is then a straightforward algebraic exercise to show the field equation of the ten-dimensional axion in the form
\begin{equation}
    dF_{(9)}= 0 \,.
\end{equation}
Let us also give the combination
\begin{eqnarray}
F_{(1)}\wedge F_{(9)} &=&
-  \mathrm{g}^2\,e^{8\phi}\,
\Delta^{-3}
\,U^2\, \omega_{10}
+ {\rm e}^{-2} \, \Delta^{-3}\,
(\partial_t \Delta)^2 \, \omega_{10}
\\
&&{}
- 
{\rm e}^{-2} \,  \Delta^{-2}\,
\mu_k\,\mu_p\,(T^{-1})^{mn}\,D_t T_{np}\, D_tT_{km}\,\omega_{10}  \,,
\nonumber
\end{eqnarray}
with
\begin{equation}
\omega_{10} = 
\frac1{9!\,{\rm g}^9}\,
\varepsilon^{i_1 \dots i_{10}}\,
{\rm e} \,  \Delta\,
dt \wedge D \mu_{i_1}\wedge\ldots\wedge D\mu_{i_{9}}\,\mu_{i_{10}} 
    \,,
\end{equation}
which shows up in the ten-dimensional dilaton equation. This equation may be shown --- analogously to the axion equation, and similar to other dimensions \cite{Cvetic:2000dm,Ciceri:2023bul}
 --- to hold as a consequence of the one-dimensional field equations. The following identity proves useful
\bea
{ *({T}_{ij}\, \mu^i\, {\cal D}\mu^j)} = \frac{\mathrm{e}}{8!\,\mathrm{g}^7} \,e^{-8\phi}\,\epsilon_{i_1 \dots i_{10}}\, 
 T_{i\ell}\, \mu^i\, ( \Delta\,  T_{i_1 \ell} -
 T_{i_1 j}\,  T_{k \ell}\, \mu^j\, \mu^k) \, \mu^{i_2}\,
{\cal D}\mu^{i_3}\wedge \dots \wedge {\cal D}\mu^{i_{10}}\wedge  dt 
\,.\label{idF1}
\eea

As for the ten-dimensional Einstein equations, we follow the tradition set in \cite{Cvetic:2000dm}, and have not explicitly checked their reduction. Their consistency has been confirmed in all the explicit solutions that we have examined. Moreover, the ansatz (\ref{eq:Ansatz}) is consistent with an extrapolation of all previously-established special cases. Let us finally point out, that the reduction ansatz (\ref{eq:Ansatz}) can be generalized to the reduction of the Lagrangian (\ref{eq:LEu}) in $D$ dimensions, on an $S^{D-1}$ sphere, again down to a one-dimensional theory. The general formulas have been given in \cite{Ciceri:2025maa}.

\subsection{One-dimensional theory}

Let us now discuss in more detail the Lagrangian (\ref{eq:L1Eu}) of the one-dimensional theory.
The 54 scalar fields parametrizing the symmetric matrix $T_{ij}$ are governed by a gauged coset space sigma model $G/K={\rm SL}(10)/{\rm SO}(10)$.
To describe their couplings to the fermionic fields later on, it is useful to introduce the coset representative ${V}_i{}^a$, defined by
\begin{equation}
    T_{ij} = {V}_i{}^a {V}_j{}^a 
    \,,
\end{equation}
up to local ${\rm SO}(10)_K$ gauge freedom
\begin{equation}
    \delta {V}_i{}^a = {V}_i{}^b\,h^{ba}\,,\qquad h \in \mathfrak{so}(10)_K
    \,.
    \label{eq:SO10K}
\end{equation}
We have introduced a subscript in the denominator group of the coset space ${\rm SO}(10)_K$ in order to distinguish it from the ${\rm SO}(10)_{\rm g}$ gauge group, under which the fields transform as
\begin{equation}
    \delta_\lambda V_i{}^a =\mathrm{g}\, \lambda_{ji} V_j{}^a\,,\quad\delta_\lambda T_{ij}= 2\,\mathrm{g}\,\lambda_{k(i}T_{j)k}\,,
    \quad \delta_\Lambda A_{ij}=D_t \lambda_{ij}
    \,.
    \label{eq:SO10G}
\end{equation}
This local ${\rm SO}(10)_{\rm g}$ symmetry has its origin in the isometry group of the sphere $S^9$.
{}From the coset representative ${V}_i{}^a$, one builds the (gauge-covariant) Maurer–Cartan current,
\begin{align}
J^{ab} = (V^{-1})^{a}{}^{i} D_t V_{i}{}^b&=(V^{-1})^{a}{}^{i}  \left(\dot V_{i}{}^b-\mathrm g\,A_{ji}V_{j}{}^b\right)\;\;\in\;\mathfrak{sl}(10)\,,
\end{align}
which decomposes into
\begin{equation}
P^{ab}=J^{(ab)}\,,\qquad
{Q}^{ab}=J^{[ab]}\;\in\;\mathfrak{so}(10)_K\,.
\end{equation}
In particular, the scalar kinetic term can be written as
\begin{equation}
    \frac{1}{4}\, 
({T}^{-1})^{ik}({T}^{-1})^{jl}\,
D_t{T}_{kl}\,  D_t\,{T}_{ij} = P^{ab} P^{ab}
\,.
\end{equation}
The scalar potential for the $T_{ij}$ in (\ref{eq:L1Eu}) is of the form typical for the truncations on spheres \cite{Nastase:2000tu,Cvetic:2000dm}.

The 45 vector fields $A_{ij}=-A_{ji}$ do not have a kinetic term in one dimension, and contribute a negative degree of freedom. They appear only within the covariant derivatives of the Lagrangian (\ref{eq:L1Eu}) and act as Lagrange multipliers that give rise to the last equation of (\ref{eq:eomEOM}).
All gravitational couplings are carried by the einbein ${\rm e}$, such that the action comes with one-dimensional diffeomorphism symmetry
\begin{equation}
    t\rightarrow \tilde t\,,\qquad
    {\rm e} \rightarrow {\rm e}/ \partial_t\tilde{t}
    \,.
    \label{eq:diff}
\end{equation}
There is no Einstein-Hilbert term in one dimension, and the graviton accounts for a negative degree of freedom, as manifest by the constraint that is obtained by variation of (\ref{eq:L1Eu}) w.r.t.\ the einbein~${\rm e}$. In counting, this compensates the degree of freedom carried by the dilaton field $\phi$.
In total, the Lagrangian (\ref{eq:L1Eu}) thus realizes $54-45$ of the degrees of freedom of the BPS multiplet (\ref{eq:BPS2}).

The gauge symmetry (\ref{eq:SO10G}) can be fixed by bringing $T_{ij}$ into diagonal form $T_{ij}=\delta_{ij}\,e^{\varphi_j}$, with $\sum_{i}\varphi_i=0$. In this gauge, the field equations (\ref{eq:eomEOM}) take the form
\begin{align}
40\,\dot\phi^2
  =\,&\sum_i \dot\varphi_i^2
  -2\,{\rm g}^2\,\mathrm{e}^{2}\,e^{8\phi}\,V_0\,,\nonumber\\
5\,\partial_t(\mathrm{e}^{-1}\dot\phi)
=\,&-{\rm g}^2\,\mathrm{e}\,e^{8\phi}\,V_0\,,\nonumber\\[1ex]
\partial_t\big( \mathrm{e}^{-1}(\dot\varphi_i -\dot\phi) \big)
  =\,&
2\,{\rm g}^2\,\mathrm{e}\,e^{8\phi}\,
\Big(
2\,e^{2\varphi_i}- \sum_k e^{\varphi_i+\varphi_k}
\Big)
 \,,
 \label{eq:EOMdiag}
 \end{align}
with 
\begin{equation}
V_0 = 
2\,\sum_k e^{2\varphi_k} - \Big(\sum_k e^{\varphi_k}\Big)^2 
\,.
\end{equation}
Moreover, the gauge fields $A_{ij}$ are set to zero by their own field equations, or decouple from the dynamics.

The field equations (\ref{eq:EOMdiag}) admit a distinguished ${\rm SO}(10)$ invariant solution, given by
\begin{equation}
\varphi_{i}=0 
\,\Leftrightarrow\,T_{ij}=\delta_{ij}
\,,\quad
e^\phi=\mathrm{g}^2\,t^2\,,\quad
{\rm e} =\mathrm{g}^{-9}\,t^{-9}\,.
\label{eq:solutionso10}
\end{equation}
Plugging this into the reduction ansatz (\ref{eq:Ansatz}) precisely yields the D$(-1)$ instanton solution (\ref{eq:D-1brane}) of Euclidean IIB supergravity, with $\mathrm{h}=0$ and $\mathrm{Q}=\mathrm{g}^{-8}$. We will show below, that this solution is also half-supersymmetric in the one-dimensional theory and actually belongs to a larger class of half-supersymmetric solutions.

%%%%%%%%%%%%%%%%%%%%%%%%
\subsection{Uplift to 12D}
%%%%%%%%%%%%%%%%%%%%%%%%

Following \cite{Tseytlin:1996ne}, the 
Euclidean axio-dilaton gravity model \eqref{eq:LEu} can be obtained by dimensional reduction of pure Einstein gravity in twelve dimensions with signature $(1,11)$ on a $(1,1)$-torus. The truncation Ansatz for the Lorentzian metric in twelve dimensions is
\bea
ds_{12}^2=-e^{-\Phi} d\tau^2+e^{\Phi}\big(dy+(c+\mathcal X)\,d\tau\big)^2+ds_{10}^2\,,\label{eq:12Duplift}
\eea
where $\tau$ and $y$ denote respectively the time and space coordinates of the $(1,1)$-torus, and $c$ is an integration constant. The D$(-1)$ instanton solution \eqref{eq:D-1brane} with $h=0$ uplifts to the following ${\rm SO}(10)$-invariant pp-wave
\bea
ds_{12}^2=-H^{-1}d\tau^2+dy^2+H\big(dy+(c+H^{-1})\,d\tau)^2+dt^2+t^2d\Omega_9^2\,,
\eea
with $H(t)=\frac{Q}{t^8}$. For $c<0$, the metric can be written in the standard pp-wave form
\bea
ds_{12}^2=dudv+f(t)\,du^2+dt^2+t^2d\Omega_9^2\,,
\eea
with
\bea
u=\sqrt{-c}\;\Big(\frac1cdy-d\tau\Big)\,,\;\;\;\;v=\sqrt{-c}\;\Big(\frac1cdy+d\tau\Big)\,,\;\;\;\;f(t)=-1-cH\,.
\eea

\subsection{Noncompact gauge groups}

The previous construction can be generalized to compactifications in which the sphere $S^9$ is replaced by a hyperboloid. In this case, the resulting one-dimensional theory (\ref{eq:L1Eu}) features a non-compact or even non-semisimple gauge group similar to the analogous constructions in higher dimensions \cite{Hull:1988jw,Hohm:2014qga}.
This generalization is straightforwardly obtained from the above results upon the introduction of a symmetric constant embedding tensor $\theta^{ij}$, by means of which the reduction formulas (\ref{eq:Ansatz}), (\ref{eq:covD}), as well as the one-dimensional Lagrangian (\ref{eq:L1Eu}) are recast in a form that is formally invariant under an ${\rm SL}(10)$ action on the indices $i, j, \dots$, i.e.
\begin{equation}
    \delta T_{ij} = 2\,\kappa_{(i}{}^k T_{j)k}\,,\quad
    \delta A_{ij} = -2\,\kappa_{[i}{}^k A_{j]k}\,,\quad
    \delta \theta^{ij} = - 2\,\kappa_k{}^{(i}\theta^{j)k}\,,\quad
    \delta \mu_i = \kappa_i{}^j \mu_j
    \,.
    \label{eq:SL10}
\end{equation}

For example, the reduction ansatz takes the following explicit form
\begin{align}
d s_{10}^2 =\,&  e^{9\phi}\,\Delta\, \mathrm{e}^2\,dt^2 
+ {\rm g}^{-2}\, e^{\phi} \, ({T}^{-1})^{ij}\, 
{ D}\mu_i\, { D}\mu_j
\,,\nonumber\\[3mm]
e^ {\Phi} =\,&  
e^{-4\phi}\,\Delta^{-1}
\,,\nonumber\\
\mathcal X=&-\frac1{2\,{\rm g}\mathrm{e}}
\left(
(T^{-1})^{ij}\, D_t T_{kj}\,\mu_{i}\mu_{l}\,\theta^{kl}-\dot\phi\right)\,.\label{eq:AnsatzTheta}
\end{align}
where the embedding coordinates $\mu_i$ are now defined by 
\begin{equation}
    \mu_i \mu_j\,\theta^{ij} = 1\,,
    \label{eq:hyper}
\end{equation}
defining a hyperboloid.
Covariant derivatives are defined as
\begin{align}
{ D} \mu_i =\,& d\mu_i - {\rm g}\, A_{ji}\,\theta^{kj}\,\mu_k\,dt\,,
\nonumber\\
 { D}_t  T_{ij} =\,&
  \dot{T}_{ij} -2\, {\rm g}\,A_{k(i}  \,\theta^{kl}\, T_{j)l}\,,
\label{eq:covDTheta}
\end{align}
and the warp factor $\Delta$ is given by
\begin{equation}
\Delta=T_{ij}\,\theta^{ik}\theta^{jl} \mu_k\mu_l
\,.
\label{eq:DeltaTheta}
\end{equation}
Consistency of this reduction ansatz then directly follows from covariantization of the previous results. The resulting one-dimensional theory is given by
\begin{align}
\mathcal L_1 =\,& 10\,\mathrm{e}^{-1}\, \dot \phi^2- \mathrm{e}^{-1}\, 
P^{ab}P^{ab}
-\frac12\,\mathrm{e}\, {\rm g}^2 \,e^{8\phi}\left(
2\,\theta^{ik}\theta^{jl}\,T_{ij}T_{kl}-( \theta^{ij}T_{ij})^2\right)\,,
\label{eq:L1EuTheta}
\end{align}
likewise invariant unde (\ref{eq:SL10}).
For $\theta^{ij}=\delta^{ij}$, the construction reproduces the above sphere compactification with gauge group ${\rm SO}(10)$. In the general case, without loss of generality, we may assume $\theta^{ij}$ to be of diagonal form with eigenvalues in $\{0,\pm1\}$.
With
\begin{equation}
    \theta^{ij}
    ={\rm diag}(\underbrace{+1, \dots, +1}_{p}, \underbrace{-1, \dots, -1}_q, \underbrace{0, \dots, 0}_r)^{ij}
    \,,
    \label{eq:thetapqr}
\end{equation}
the gauge group of the one-dimensional theory is the group ${\rm CSO}(p,q,r)_\mathrm{g}$. For $r=0$, this is the non-compact group ${\rm SO}(p,q)_\mathrm{g}$\,. Note that for any choice of $\theta^{ij}$, the ten-dimensional metric (\ref{eq:AnsatzTheta}) remains Euclidean.
With (\ref{eq:thetapqr}), the internal space (\ref{eq:hyper}) is a product of $r$ noncompact directions and a hyperboloid that following \cite{Hull:1988jw} we denote by $H^{p,q}$.
Let us also point out that for generic values of $p$, $q$, $r$, setting the matrix of scalar fields to $T_{ij}=\delta_{ij}$, no longer describes a solution of the theory (\ref{eq:L1EuTheta}). Consequently, in this case, (\ref{eq:AnsatzTheta}) with $T_{ij}=\delta_{ij}$ is no longer a solution of the Euclidean IIB theory. Yet, (\ref{eq:AnsatzTheta}) continues to describe a consistent truncation of the IIB theory.

\subsection{Spacetime signatures}

So far, we have described the consistent truncation of the Euclidean theory (\ref{eq:LEu}). The same truncation procedure may be extended to theories defined with different spacetime signatures.

Let us first consider the standard Lorentzian IIB supergravity, which contains the dilaton-axion subsector
\begin{equation}
{\cal L}_{10, \rm{Lor}} =
 |E|\Big(R - \tfrac12 \partial_\mu \Phi \,\partial^\mu\Phi  
-\tfrac12 e^{2\Phi}\, \partial_\mu \mathcal X \,\partial^\mu \mathcal X \Big) \,,
\label{eq:LLo}
\end{equation}
in Lorentzian spacetime. This differs from (\ref{eq:LEu}) by the sign in front of the axion kinetic term. The former truncation ansatz (\ref{eq:AnsatzTheta}) can be adapted by formally sending $\mathrm{e}\rightarrow i \mathrm{e}$\footnote{Equivalently, one could rescale $\mathrm{g}\rightarrow i\mathrm{g}$, $A_{ij}\rightarrow -iA_{ij}$ and flip the sign of the ten-dimensional metric.}
\begin{align}
d s_{10}^2 =\,&  - e^{9\phi}\,\Delta\, \mathrm{e}^2\,dt^2 
+ {\rm g}^{-2}\, e^{\phi} \, ({T}^{-1})^{ij}\, 
{ D}\mu_i\, { D}\mu_j
\,,\nonumber\\[3mm]
e^ {\Phi} =\,&  
e^{-4\phi}\,\Delta^{-1}
\,,\nonumber\\
\mathcal X=&-\frac1{2\,{\rm g}\mathrm{e}}
\left(
(T^{-1})^{ij}\, D_t T_{kj}\,\mu_{i}\mu_{l}\,\theta^{kl}-\dot\phi\right)\,,\label{eq:AnsatzLor}
\end{align}
such that the ten-dimensional metric is of signature $(1,9)$, and an imaginary factor has been absorbed into the axion in order to achieve the sign flip in (\ref{eq:LLo}). Accordingly, the one-dimensional Lagrangian is given by
\begin{align}
\mathcal L_{1, \mathrm{Lor}} =\,& 10\,\mathrm{e}^{-1}\, \dot \phi^2-\mathrm{e}^{-1}\,P^{ab}P^{ab}
+\frac12\,\mathrm{e}\, {\rm g}^2 \,e^{8\phi}\left(
2\,\theta^{ik}\theta^{jl}\,T_{ij}T_{kl}-( \theta^{ij}T_{ij})^2\right)\,,
\label{eq:L1Lo}
\end{align}
which differs from (\ref{eq:L1EuTheta}) by the sign in front of the potential term. 
The gauge group of the one-dimensional theory is still determined by the choice of the embedding tensor $\theta^{ij}$, and given by ${\rm CSO}(p,q,r)_\mathrm{g}$ for a $\theta^{ij}$ of the form (\ref{eq:thetapqr}). In particular, for $\theta^{ij}=\delta^{ij}$, the theory still possesses the  local ${\rm SO}(10)_\mathrm{g}$ gauge symmetry (\ref{eq:SO10G}) in accordance with the isometry group of $S^9$.
However, in this case the truncation no longer carries an ${\rm SO}(10)$ invariant solution $T_{ij}=\delta_{ij}$. This is a consequence of the signs in the field equation for the einbein $\mathrm{e}$
\begin{equation}
    10 \,\dot \phi^2 + 40\,\mathrm{e}^2\, {\rm g}^2 \,e^{8\phi} = 0 \,.
\end{equation}
In contrast, for $\theta^{ij}={\rm diag}(1,1,1,1,1,-1,-1,-1,-1,-1)$, i.e.\ for $p=q=5$, the ${\rm SO}(5,5)$ gauged theory (\ref{eq:L1Lo}) admits a solution with constant scalars $T_{ij}=\delta_{ij}$, running dilaton $e^\phi=\mathrm{g}^2\,t^2$, and ${\rm e} =\frac12\,\mathrm{g}^{-9}\,t^{-9}$. Its uplift (\ref{eq:AnsatzLor}) corresponds to a solution of Lorentzian IIB supergravity, whose internal space is a Euclidean hyperboloid $H^{5,5}$\,.

Finally, we will also be interested in embedding the Lagrangian (\ref{eq:LEu}) with its `wrong' sign in front of the kinetic axion term into spacetimes of different signature. As mentioned above, this includes the exotic IIB' and IIB$^*$ theories \cite{Hull:1998vg,Hull:1998ym}, of spacetime signature $(9,1)$ and $(5,5)$, in which all fields can be taken real, while some kinetic terms, in particular that of the axion, acquire opposite signs.
In this case, the internal space in the reduction to one dimension is of indefinite signature, which results in a scalar target space of indefinite signature. Specifically, the reduction of (\ref{eq:LEu}) in a ten-dimensional spacetime of signature $(s,t)$ yields a one-dimensional theory (\ref{eq:L1EuTheta}) with the coset space ${\rm SL}(10)/{{\rm SO}(10)}$ replaced by a different real form ${\rm SL}(10)/{{\rm SO}(s,t)_K}$. The matrix $T_{ij}$ is then realized as
\begin{equation}
    T_{ij} = {V}_i{}^a {V}_j{}^b\,\eta_{ab} 
    \,,
\end{equation}
in terms of the coset representative $V_i{}^a$, with the symmetric ${\rm SO}(s,t)_K$ invariant tensor $\eta_{ab}$.
The Maurer–Cartan currents are built as
\begin{align}
J^{ab} = \eta^{ac}(V^{-1})_c{}^{i} D_t V_{i}{}^b&=\eta^{ac}(V^{-1})_c{}^{i} \left(\dot V_{i}{}^b-\mathrm g\,A_{ji}\theta^{jk} V_{k}{}^b\right)\;\;\in\;\mathfrak{sl}(10)\,,
\end{align}
which decomposes into
\begin{equation}
P^{ab}=J^{(ab)}\,,\qquad
{Q}^{ab}=J^{[ab]}\;\in\;\mathfrak{so}(s,t)_K\,.\label{eq:PandQ}
\end{equation}
The reduction ansatz remains identical (\ref{eq:AnsatzTheta}), and the resulting one-dimensional theory is given as
\begin{align}
\mathcal L_{1} =\,& 10\,\mathrm{e}^{-1}\, \dot \phi^2-\mathrm{e}^{-1}\,
P^{ab}P^{cd}\,\eta_{ac}\eta_{bd}
-\frac12\,\mathrm{e}\, {\rm g}^2 \,e^{8\phi}\left(
2\,\theta^{ik}\theta^{jl}\,T_{ij}T_{kl}-( \theta^{ij}T_{ij})^2\right)\,.
\label{eq:L1All}
\end{align}

To summarize, the one-dimensional Lagrangian (\ref{eq:L1All}) depends on the choice of the two tensors $\eta^{ab}$ and $\theta^{ij}$. The first one defines the scalar coset space and captures the signature of the ten-dimensional spacetime, whereas the latter defines the gauge group of the one-dimensional theory and defines the shape and isometries of the internal space.
For $\eta=\theta$, i.e.\ $p=s$ and $q=t$, the field equations of (\ref{eq:L1All}) admit a solution analogous to (\ref{eq:solutionso10})
\begin{equation}
V_i{}^a=\delta_i{}^a \,\Leftrightarrow\,(T^{-1})^{ij}=\theta^{ij}
\,,\quad
e^\phi=\mathrm{g}^2\,t^2\,,\quad
{\rm e} =\mathrm{g}^{-9}\,t^{-9}\,.
\label{eq:solutionSignature}
\end{equation}
The corresponding ten-dimensional solution (\ref{eq:AnsatzTheta}) still features a flat metric, but now of signature $(p,q)$. For $p=9$, $q=1$, i.e.\ Lorentzian ten-dimensional spacetime and gauge group ${\rm SO}(9,1)$, the solution (\ref{eq:solutionSignature}) corresponds to the solution of IIB$^*$ supergravity, considered in \cite{Blair:2025nno}.

%%%%%%%%%%%%%%%%%%%%%%%%%%%%%%%%%%%%%%%%%%%%%%%

\section{Maximal supergravities in one dimension}\label{sec:SUGRA}
In this section, we construct the maximally supersymmetric extension of the one-dimensional Lagrangian \eqref{eq:L1All}, allowing for an arbitrary $\theta^{ij}$, while restricting $\eta^{ab}$ to have signature $(10,0)$ or $(9,1)$. The resulting $\mathrm{CSO}(p,q,r)_\mathrm{g}$-gauged supergravities\footnote{Although there is no notion of propagating gravitational degrees of freedom in one dimension, we still refer to the one-dimensional theory as a supergravity theory. We emphasize, however, that its Lagrangian encodes the dynamics of the lowest fluctuations of a genuine gravitational theory in ten dimensions.} are expected to describe Kaluza-Klein truncations of the Euclidean type IIB supergravity and of the Lorentzian type IIB'/IIB${}^*$ supergravities, respectively. As explained in \cite{Bergshoeff:2007cg,DHoker:2025nid}, the former should be treated as a complex theory, while the latter correspond to real sections  distinguished by different reality conditions of the fields. 
For a uniform discussion of all cases, we treat all one-dimensional fields and symmetry parameters as complex in this section. This is possible because the supersymmetric Lagrangian does not involve complex conjugation. The dependence of this Lagrangian on the fields is said to be holomorphic, and the corresponding supersymmetry transformations are viewed as holomorphic transformations on a complex space \cite{Nicolai:1978vc}.\footnote{Strictly speaking, the one-dimensional theory constructed in this section should therefore be viewed as a consistent truncation of complex type IIB${}_\mathbb{C}$ supergravity \cite{Bergshoeff:2007cg,DHoker:2025nid}.} The real forms of the theory, available in signature $(9,1)$, will be discussed in Section~\ref{sec:realforms}. Let us stress that throughout this work, our main interest lies in the bulk realization of the Euclidean IKKT model, which is based on one-dimensional supergravity with $\mathrm{CSO}(p,q,r)_{\mathrm{g}}=\mathrm{SO}(10)_\mathrm g$ and $\mathrm{SO}(s,t)_K=\mathrm{SO}(10)_K$.%\footnote{Euclidean IIB supergravity should be viewed as a particular slice of the complex IIB$_{\mathbb{C}}$ theory, in which the metric is real in order to have a notion of (10,0) spacetime signature. In this case, however, the reality conditions are not preserved by supersymmetry or Lorentz transformations and the action remains complex. The situation is the analogous for the one-dimensional theory describing its Kaluza-Klein truncations. It must be thought of as a slice where, in particular, the scalar matrix $T_{ij}$ real in order to have a notion of $\mathrm{SO}(10)_K$.}

In addition to the fermionic fields which will be discussed shortly, maximal supersymmetry also necessitates enlarging the bosonic sector by adding 120 axionic scalar fields $a_{ijk}=a_{[ijk]}$. This is in accordance with the representations of the BPS multiplet \eqref{eq:BPS2} in the IKKT model. In the following, we will refer to these scalars simply as `axions'; however, they should not be confused with the type IIB supergravity axion field $\mathcal X$ in ten dimensions. In particular, the latter is non-zero (\ref{eq:AnsatzTheta}) even for vanishing $a_{ijk}$. Non-vanishing $a_{ijk}$ will rather trigger p-form gauge fields in ten dimensions. For later convenience, we introduce the $\mathrm{SO}(s,t)_K$ covariant combinations of scalar fields
\begin{align}
 \mathcal{T}^{ab}&:=\theta^{ij}\,V_i{}^aV_j{}^b\,,\nonumber\\
a_{abc}&:=V_{abc}{}^{ijk}\,a_{ijk}\,,\label{eq:axionSO}
\end{align} 
where in this, and in the following, we abbreviate $(V^{-1})_a{}^i=:V_a{}^i$, and use the notation 
\begin{equation}
    V_{a_1\ldots a_n}{}^{i_1\ldots i_n}:=V_{a_1}{}^{[i_1}\ldots V_{a_n}{}^{i_n]}\,. 
\end{equation}
Note that from now on we also lower and raise the indices $a\,,b=1\,,\ldots10,$ using implicitly the $\mathrm{SO}(s,t)_K$ invariant tensor $\eta_{ab}$ and its inverse $\eta^{ab}$. We have in particular,
\begin{equation}
V_a{}^{i}V_{ib}=\eta_{ab}\,.
\end{equation}

Due to the triviality of the tangent space in one dimension, the fermionic fields of the theory consist of Grassmann variables that transform in different $\mathrm{SO}(s,t)_K$ representations. These include the gravitino $\psi^\alpha$ and a dilatino $\lambda^\alpha$, which both transform as spinors under $\mathrm{SO}(s,t)_K$, as well as the vector-spinor $\chi^\alpha_a$. Here, we employ a generalization of the spinor notations introduced in \eqref{eq:IKKTaction} for the specific case of $\mathrm{SO}(10)_\mathrm g$ and $\mathrm{SO}(10)_K$. In particular, we define charge-conjugated spinors using a symmetric charge conjugation matrix $\mathcal C$, such that we have for instance,
\begin{equation}
\bar\psi_\alpha:=\psi^\beta\,\mathcal C_{\alpha\beta}\,.
\end{equation}
 It is important to note, however, that unlike the chiral spinor $\Psi^\alpha$ in the IKKT model, a supergravity spinor such as the gravitino $\psi^\alpha$ (or the supersymmetry parameter $\epsilon^\alpha$) carries here 32 independent components. Likewise, the vector-spinor $\chi^\alpha_a$ satisfies the trace condition
\begin{equation}
     (\Gamma^a)^\alpha{}_\beta\,\chi^\beta_a=0\,,\label{eq:tracechi}
\end{equation}
and carries $2\times 144$ components. The $\mathrm{SO}(s,t)_K$ gamma matrices satisfy 
\begin{equation}
    \{\Gamma_a,\Gamma_b\}=2\,\eta_{ab}\,\mathbb I_{32}\,.
\end{equation} 
Their symmetry properties are summarized in Appendix~\ref{app:gamma}. As mentioned previously, the reality conditions that can be imposed on the spinors depend on the signature of $\eta_{ab}$ and will be discussed later on. For the time being, all spinors components are treated as complex.

The couplings of the scalars to the fermionic fields involve the $\mathrm{SO}(s,t)_K$ covariant projection of the Maurer-Cartan current $P^{ab}$, defined in \eqref{eq:PandQ}, as well as the dressed current for the axions
\begin{equation}
    p_{abc}:=
     V_{abc}{}^{ijk}\,
    D_t a_{ijk}=V_{abc}{}^{ijk}\,\big( \dot a_{ijk}-3\,\mathrm{g}\,\theta^{lm}A_{l[i}\,a_{jk]m}\big)\,.
\end{equation}
The second component $Q^{ab}$ of the Maurer-Cartan current \eqref{eq:PandQ} is a composite connection which enters the definition of the $\mathrm{SO}(s,t)_K$ covariant derivative $\mathcal D_t$. The latter acts on the fermionic fields as, 
\begin{align}
\mathcal{D}_t \psi^\alpha&=\dot\psi^\alpha+\frac14 Q^{ab}\,(\Gamma_{ab})^{\alpha}{}_{\beta}\,\psi^\beta\,,\nonumber\\
\mathcal D_t\chi^{\alpha}_a&=\dot\chi^{\alpha}_a+\frac14 Q^{bc}\,(\Gamma_{bc})^{\alpha}{}_{\beta}\,\chi^{\beta}_a+ Q^{ab}\,\chi^{\alpha}_b\,,
\end{align} 
and analogously on $\lambda^\alpha$. Note that we have so far kept spinor indices explicit, but in the following they will be often suppressed to avoid cluttering.

\subsection{Lagrangian and supersymmetry transformations}
The Lagrangian of gauged maximal supergravity, up to second order in the fermionic fields, can be decomposed as follows
\begin{equation}
\mathcal L_{\text{SUGRA}}=\mathcal L_{\text{0}}+\mathrm{g}\,\mathcal L_{\text{top}}+\mathrm g\,\mathcal L_{\text{Yukawa}}+\mathrm{g}^2\mathcal L_{\text{pot}}\,.\label{eq:LSUGRA}
\end{equation}
The first term depends on the gauge coupling constant only implicitly through the covariant derivatives and currents. Its expression is given by
\begin{align}
\mathcal L_0=&\,10\,\mathrm{e}^{-1} \dot \phi^2-\mathrm{e}^{-1}P^{ab} P_{ab}
-\frac1{12}\,\mathrm{e}^{-1}e^{-2\phi}\,p_{abc}\,p^{abc}
+20\,\bar\lambda\,\mathcal D_t\lambda+2\,\bar\chi^a\,\mathcal D_t\chi_a
\nonumber\\&{}
-20\,\mathrm{e}^{-1}\bar\psi\,\Gamma_*\lambda\,\dot\phi
+2\,\mathrm{e}^{-1}\bar\psi\,\Gamma_{b}\chi_{a}\, P^{ab}-\frac{1}{2}\mathrm{e}^{-1}e^{-\phi}\,\bar\chi_{a}\,\Gamma_{bc}\,\psi\,p^{abc}
-\frac{1}{12}e^{-\phi}\,\bar\chi^{a}\,\Gamma_{bcd}\,\chi_a\,p^{bcd}\nonumber\\
\,&{}
-e^{-\phi}\,\bar\chi_a\,\Gamma_b\,\chi_c\,p^{abc}-\frac{1}{6}\mathrm{e}^{-1}e^{-\phi}\,\bar\lambda\,\Gamma_{abc}\,\Gamma_*\psi\,p^{abc}-e^{-\phi}\,\bar\lambda\,\Gamma_{abc}\lambda\,p^{abc}-e^{-\phi}\,\bar \chi_c\,\Gamma_{ab}\Gamma_*\,\lambda\,p^{abc}\,.
\end{align}
where the first line comprises the kinetic terms for the various fields, while the rest consists of Noether-type couplings between the fermions and the scalar currents. Note that there is no kinetic term for the gravitino, as the standard Rarita-Schwinger term vanishes in one-dimension. The various terms in $\mathcal L_0$ are determined in parallel with the supersymmetry transformations of the fields, by starting from the bosonic kinetic terms \eqref{eq:L1All} and imposing supersymmetry at zeroth order in $\mathrm{g}$.\footnote{The initial Ansatz for the supersymmetry variations is strongly constrained by the symmetry properties of the fields and the gamma matrices, as well as by the requirement that they close according to a standard gauged supergravity algebra, cf.~\ref{sec:SUSYalgebra}.}  For the bosonic fields, the supersymmetry variations read
\begin{align}
\delta_\epsilon \mathrm{e}&=\bar\epsilon\, \psi\,,\nonumber\\[1mm]
\delta_\epsilon \phi&=\bar\epsilon\,\Gamma_*\lambda\,,\nonumber
\\[1mm]
\delta_\epsilon V_i{}^a&=V_i{}_b\,\,\bar\epsilon\,\Gamma^{(a}\,\chi^{b)}\,,\nonumber\\[1mm]
\delta_\epsilon a_{ijk}&=-e^\phi\,V_{ijk}{}^{abc}\left(
3\,\bar\epsilon\,\Gamma_{ab}\,\chi_{c}
+\bar\epsilon\,\Gamma_{abc}\Gamma_*\,\lambda\right)\,,
\label{eq:SUSYbos}
\end{align}
while those of the fermions take the form
\begin{align}
\delta_\epsilon \psi&= \mathcal D_t\,\epsilon
+\frac1{24}e^{-\phi}\,p^{abc}\,\Gamma_{abc}\,\epsilon \,
+\mathrm{e} \,\mathrm{g}\,A\,\epsilon\,,\nonumber\\[1mm]
\delta_\epsilon\lambda&=\frac{1}{2}\mathrm{e}^{-1}\dot\phi\,\Gamma_{*}\,\epsilon
+\frac{1}{240}\mathrm{e}^{-1}e^{-\phi}\,p^{abc}\,\Gamma_{abc}\Gamma_*\,\epsilon 
+\mathrm g\,B\,\epsilon\,,\nonumber\\[1mm]
\delta_\epsilon\chi^{a}&=\frac{1}{2}\mathrm{e}^{-1}P^{ab}\,\Gamma_{b}\,\epsilon
+\frac{7}{80}\mathrm{e}^{-1}\,e^{-\phi}\,p_{bcd}\,\tilde\Gamma^{ab,cd}\,\epsilon
+\mathrm g\,C^a\,\epsilon\,,\label{eq:SUSYall}
\end{align}
where in the last variation we have introduced the gamma matrix combinations
\begin{equation}
\tilde\Gamma^{a_1a_2\,,\, a_3\ldots a_n}:=\eta^{a_1a_2}\,\Gamma^{a_3\ldots a_n}-\frac{1}{11-n}\Gamma^{a_1\ldots a_n}\,\label{eq:Gammacombi}
\end{equation}
for $n>2$, which satisfy $\Gamma_{a_1}\tilde\Gamma^{a_1[a_2\,,\,a_3\ldots a_n]}=0$. The tensors $A,B,C^a$ appearing in the variations \eqref{eq:SUSYall} are known as the \textit{fermion shifts}. They depend on the scalar fields and gamma matrices and, as such, carry spinor indices which we have suppressed. Their expressions, together with the variation of the gauge field $A_{ij}$, are determined in a second step by imposing supersymmetry of the Lagrangian at order $\mathrm{g}$.

The second term in the Lagrangian \eqref{eq:LSUGRA} corresponds to a purely bosonic `topological' term. Its expression is independent of the einbein and involves a single derivative,
\begin{align}
    \mathcal L_{\text{top}}=&\,\frac{\mathrm{i}^{(s-1)}}{1152}\,\varepsilon_{a_1a_2a_3a_4a_5a_6a_7a_8a_9 a_{10}}\,\mathcal{T}_{bc}\,a^{ba_1a_2}\,a^{ca_3a_4}\,a^{a_5a_6a_7}\,p^{a_8a_9a_{10}}\,,\label{eq:Ltop}
\end{align}
with $\varepsilon_{a_1\ldots a_{10}}$ the $\mathrm{SO}(s,t)_K$ Levi-Civita tensor. Due to the power of $\mathrm{i}$, this is the only term in the Lagrangian that depends explicitly on the $(s,t)$ signature. 
The third term in \eqref{eq:LSUGRA} describes Yukawa-type couplings between fermions and scalars. It decomposes into the following spinor bilinears,
\begin{equation}
\mathcal L_{\text{Yukawa}}=-40\,\bar\lambda\,B\,\psi-4\,\bar\chi_a\,C^a\,\psi+\mathrm{e}\,\bar\lambda\,E\,\lambda+\mathrm{e}\,\bar\lambda\,E^a\chi_a+\mathrm{e}\,\chi_a\,E^{ab}\chi_b\,,\label{eq:YukawaAnsatz}
\end{equation}
with the new tensors $E\,,E^a\,,E^{ab}$ and the fermion shifts $B\,,C^a$ that appear in \eqref{eq:SUSYall}. Both $\mathcal L_{\text{top}}$ and $\mathcal L_{\text{Yukawa}}$, as well as the variation of the gauge field, are completely determined by requiring the supersymmetry variations of the Lagrangian to vanish at order $\mathrm{g}$. This imposes a number of linear relations between the Yukawa tensors $A,B,C^a, E, E^a, E^{ab}$ which we gather in Appendix~\ref{app:Yukawa}. As an example, we provide here the relations imposed by the vanishing of variations proportional to $\mathrm{g}\,\dot\phi\,\bar\epsilon^{\alpha}\psi^\beta$ and $\mathrm{g}\,\dot\phi\,\bar\epsilon^\alpha\chi_{a}^\beta$, 
\begin{align}
(\mathcal C\,\Gamma_*\,B)_{(\alpha\beta)}&=0\,,
\nonumber\\
8\,(\mathcal C \frac{\partial C^a}{\partial \phi})_{\alpha\beta}+(\mathcal C\Gamma_*\,E^a)_{\beta\alpha}&=0\,.
\end{align}
where $\mathcal C_{\alpha\beta}$ is the charge conjugation matrix. The full set of linear relations between the Yukawa tensors are solved to obtain 
\begin{align}
A=&-\frac14\,b\,\Gamma_*+\frac18 b^{abc}\,\Gamma_{abc}\Gamma_* -\frac{1}{32}b^{abcd}\,\Gamma_{abcd}\Gamma_* 
\,,\nonumber\\[2mm]
B=&\,\frac{b}{10}+\frac{3}{80} b^{abc}\,\Gamma_{abc} +\frac{1}{160}b^{abcd}\,\Gamma_{abcd}
\,,\nonumber\\[2mm]
C^a=&-\frac12 b^{ab}\,\Gamma_b\Gamma_*-\left(\frac{7}{80}\,b_{bcd}-\frac{1}{6}\,b_{b,cd} \right)\tilde\Gamma^{ab,cd}\Gamma_*
+\frac{3}{40}b_{bcde} \,\tilde\Gamma^{ab,cde}\Gamma_*
\,,\nonumber\\[2mm]
E=&-21\,b \, \Gamma_*-7\,b^{abc}\,\Gamma_{abc}\Gamma_*-\frac{9}{8}b^{abcd}\,\Gamma_{abcd}\Gamma_* \,,\nonumber\\[2mm]
E^a=&-16\,b^{ab}\, \Gamma_b+\left(3\,b^{abc}-4\,b^{a,bc}\right)\,\Gamma_{bc}-2\,b^{abcd}\,\Gamma_{bcd}\,,\nonumber\\[2mm]
E^{ab}=&- \left( 4\,b^{ab}-\frac1{10}\,\eta^{ab}\,b\right)\,\Gamma_*-\frac14 \eta^{ab}\,b^{cde}\,\Gamma_{cde}\Gamma_* +\frac13 \left(3\,b^{acb}-4\,b^{c,ab}\right)\,\Gamma_{c}\Gamma_* \nonumber\\[1.5mm]
&\,\,+\frac{1}{16}\eta^{ab}\,b^{cdef}\,\Gamma_{cdef}\Gamma_* 
-\frac{3}{2}b^{abcd}\,\Gamma_{cd}\Gamma_*\,,\label{eq:YukawaAll} 
\end{align}
in terms of the irreducible $\mathrm{SO}(s,t)_K$ tensors, 
\begin{align}
b&=e^{4\phi}\,\mathcal{T}^a{}_a\,,\nonumber\\
b^{ab}&=\,e^{4\phi}\left(\mathcal{T}^{ab}-\tfrac{1}{10}\,\eta^{ab}\,\mathcal{T}^c{}_c\right)\,,\nonumber\\
b^{abc}&=\,e^{3\phi}\,\mathcal{T}^{d[a}\,a^{bc]}{}_d\,,\nonumber\\
b^{a,bc}&=\,e^{3\phi}\left(\mathcal{T}^{da}\,a^{bc}{}_d-\mathcal{T}^{d[b}\,a^{c]a}{}_d\right)\,,\nonumber\\
b^{abcd}&=\,e^{2\phi}\,\mathcal{T}^{ef}\,a_e{}^{[ab}\,a^{cd]}{}_f\,,\label{eq:babcde}
\end{align}
transforming respectively in the \textbf{1}, \textbf{54}, \textbf{120}, \textbf{320} and  \textbf{210}. Note in particular that $b^{[a,bc]}=0=b^{a,ac}$. Furthermore, the gauge field variation reads
\begin{align}
\delta_\epsilon A_{ij}=&\,e^{4\phi}\,V_{ij}{}^{ab}\left(\bar\epsilon \,\Gamma_{ab}\Gamma_*\,\psi+4\,\mathrm{e}\,\bar\epsilon\,\Gamma_{ab}\,\lambda-2\,\mathrm{e}\,\bar\epsilon\,\Gamma_{a}\Gamma_*\,\chi_b\right)\nonumber\\[1mm]
&\,+e^{3\phi}\,V_{ij}{}^{bc}\,a_{abc}\left(\bar\epsilon\,\Gamma^a\Gamma_*\,\psi+3\,\mathrm{e}\,\bar\epsilon\,\Gamma^a\,\lambda+\mathrm{e}\,\bar\epsilon\,\Gamma_*\,\chi_a\right)\,.\label{eq:SUSYA}
\end{align}

The last term in the Lagrangian \eqref{eq:LSUGRA} describes the scalar potential, which contains interaction terms up to quartic order in the axion fields and can be written compactly as
\begin{align}
    \mathcal{L}_{\text{pot}}=-\mathrm{e}\,V_{\text{pot}}=-\mathrm{e}\left(b^{ab}\,b_{ab}-\frac25\,b^2
    -\frac14\,b^{abc}\,b_{abc}+\frac29\,b^{a,bc}\,b_{a,bc}+\frac3{16}\,b^{abcd}\,b_{abcd}\right)\,.  \label{eq:Lpot}
\end{align}
It is fixed by requiring supersymmetry variations of the Lagrangian to vanish at order $\mathrm{g}^2$. This requires the Yukawa tensors $A, B,\ldots$ to satisfy as set of redundant quadratic relations which we again gather in Appendix~\ref{app:Yukawa}. Finally, a more explicit form of the scalar potential is obtained by substituting the expressions for the tensors \eqref{eq:babcde},     
\begin{align}
V_{\text{pot}}=&\,\frac12\, e^{8\phi}\left(2\, \mathcal{T}^{ab}\,\mathcal{T}_{ab}-(\mathcal{T}^{a}{}_a)^2\right)-\frac14 \,e^{6\phi}
\left(2\,\mathcal{T}^{ac}\,\mathcal{T}^{bd}\,a_{abe}\,a_{cd}{}^{e}-\mathcal{T}^{ac}\,\mathcal{T}^{b}{}_{c}\,a_{ade}\,a_{b}{}^{de}\right)\nonumber\\
&\,+\frac3{16}\,e^{4\phi}\,\mathcal{T}^{ef}\,\mathcal{T}^{gh}\,a_e{}^{[ab}\,a^{cd]}{}_f\,a_{gab}\,a_{cdh}\,.
\end{align}
This expression contrasts with the scalar potential of $\mathrm{CSO}(p,q,r)_\mathrm{g}$ gauged supergravity in two spacetime dimensions \cite{Ortiz:2012ib}, which contains couplings involving up to eight axions. In the present case, it is the interplay between the variations of the topological term \eqref{eq:Ltop}  and those of the potential that ensures the iterative supersymmetric construction of the Lagrangian terminates at quartic order in the axions.

\subsection{Supersymmetry algebra}
\label{sec:SUSYalgebra}
The supersymmetry variations \eqref{eq:SUSYbos}, \eqref{eq:SUSYall} and \eqref{eq:SUSYA} close on the bosonic fields of the theory,\footnote{We do not discuss closure on the fermionic fields, which, as usual in maximal supergravity, would require using the field equations.} up to second order in fermions, into a standard gauged supergravity algebra
\begin{align}
\left[\delta_{\epsilon_1},\delta_{\epsilon_2}\right]=\delta^{\mathrm{cgct}}_\xi +\delta^{\mathrm{SO}(s,t)_K}_{ \Omega}+\delta^{\mathrm{CSO}(p,q,r)_\mathrm{g}}_{ \Lambda}\,,\label{eq:susyalgebra}
\end{align}
where $\delta^{\text{cgct}}_\xi$ denotes a covariant general coordinate (time) transformation with parameter $\xi$, combining a one-dimensional diffeomorphism \eqref{eq:diff} with gauge transformations of the form
\begin{equation}
    \hat\Omega^{ab}=-\xi\,Q^{ab}\,,\;\;\;\;\;\hat\Lambda_{ij}=-\xi\,A_{ij}\,.
\end{equation}
On the einbein and the dilaton, it simply reduces to a diffeomorphism, giving $\delta^{\mathrm{cgct}}_\xi \mathrm{e}=\partial_t(\xi \,\mathrm e)$ and $\delta_\xi \phi=\xi\dot\phi$, while on the coset representative, the axions and the gauge fields, the combined transformations lead to
\begin{equation}
\delta_\xi^{\mathrm{cgct}}V_i{}^a=\xi \,P^{ab}\,V_{ib}\,,\;\;\;\;\;\;\;\;\;\;\;\delta_\xi^{\mathrm{cgct}}a_{ijk}=\xi \,D_ta_{ijk}\,,\;\;\;\;\;\;\;\;\;\;\;\delta_\xi^{\mathrm{cgct}}A_{ij}=0\,.
\end{equation}
The parameters of the gauge transformations appearing on the r.h.s.\ of \eqref{eq:susyalgebra} are given by 
\begin{align}
\xi&=\mathrm{e}^{-1}\bar\epsilon_{2}\,\epsilon_{1}\,,\nonumber\\[1mm]
 \Omega^{ab}&=-\mathrm{g}\,e^{4\phi}\,\mathcal T^{c[a}\left(\bar\epsilon_2\,\Gamma^{b]}{}_{c}\Gamma_*\,\epsilon_1+e^{-\phi}a^{b]}{}_{cd}\,\bar\epsilon_2\,\Gamma^d\Gamma_*\,\epsilon_1\right)\,,\nonumber\\[1mm]
 \Lambda_{ij}&=e^{4\phi}\,V_{ij}{}^{bc}\left(\bar\epsilon_2\,\Gamma_{bc}\Gamma_*\,\epsilon_1+e^{-\phi}\,\bar\epsilon_2\,\Gamma^a\Gamma_*\epsilon_1\,a_{abc}\right)\,.\label{eq:gaugepara}
\end{align}

Closure on the einbein and the dilaton is straightforward to check, given that the fermion shifts among \eqref{eq:YukawaAll} satisfy
\begin{align}
(\mathcal C A)_{(\alpha\beta)}=0=(\mathcal C \,\Gamma_* B)_{(\alpha\beta)}\,, 
\end{align}
as a direct consequence of the symmetries of the gamma matrices \eqref{eq:Gammasym}. For the coset representative, successive supersymmetry variations lead to 
\begin{equation}
\left[\delta_{\epsilon_1},\delta_{\epsilon_2}\right]V_i{}^a=\xi P^{ab}\,V_{ib}+\mathrm{g} \,K^{ab}\,V_{ib}\,,\label{eq:algebraV}
\end{equation}
where $K^{ab}$ depends on the fermion shift $C^a$, and can be written as
\begin{align}
K^{ab}&=\bar\epsilon_2\,\Gamma^{(a}C^{b)}\,\epsilon_1-1\leftrightarrow 2\nonumber\\[1mm]
&=-e^{4\phi}\,\mathcal T^{c(a}\left(\bar\epsilon_2\,\Gamma^{b)}{}_c\Gamma_*\,\epsilon_1+e^{-\phi}a^{b)}{}_{cd}\,\bar\epsilon_2\,\Gamma^d\Gamma_*\,\epsilon_1\right)\nonumber\\[1mm]
&=-\Lambda_{ij}\,\theta^{jk}\,V_j{}^aV^{bi}-\mathrm{g}^{-1}\Omega^{ab}\,,
\end{align}
where the first step follows directly from using the first relation in \eqref{eq:linPpsi}. This shows that the second term in \eqref{eq:algebraV} coincides precisely with the combination of $\mathrm{CSO}(p,q,r)_\mathrm g$ and $\mathrm{SO}(s,t)_K$ on the r.h.s.\ of \eqref{eq:susyalgebra}. Applying the supersymmetry commutator of the axions leads to
\begin{align}
\left[\delta_{\epsilon_1},\delta_{\epsilon_2}\right]a_{ijk}=\xi\,D_ta_{ijk}-3\,\mathrm{g}\,e^{-\phi}V_{ijk}{}^{abc}\left(\bar\epsilon_2\,\Gamma_{ab}C_c\,\epsilon_1+\frac13\bar\epsilon_2\,\Gamma_{abc}\Gamma_* B\,\epsilon_1-1\leftrightarrow 2\right)\,.
\end{align}
By making use of the first relation in \eqref{eq:linpsi}, one readily checks that the second term indeed corresponds to the last gauge transformation on the r.h.s.\ of \eqref{eq:susyalgebra}. Finally, upon evaluating the commutator on the gauge fields, a lengthy algebra shows that the fermion shifts contributions cancel, leaving only the expected gauge transformations
\begin{equation}
\left[\delta_{\epsilon_1},\delta_{\epsilon_2}\right]A_{ij}=D_t\Lambda_{ij}\,.
\end{equation}

We recall that, so far, no reality conditions have been specified for the fields or the symmetry parameters. Real forms of the one-dimensional theory are discussed in the next section.

\subsection{Real forms in (9,1) signature}\label{sec:realforms}
We now consider specifically one-dimensional $\mathrm{CSO}(p,q,r)_\mathrm g$ gauged supergravity with $\eta_{ab}$ of (9,1) signature. In this case, we can choose a Weyl representation where all the $\mathrm{SO}(9,1)_K$ gamma matrices are real. The latter are simply obtained from $\mathrm{SO}(10)_K$ gamma matrices by rescaling $\Gamma_{10}\rightarrow -\mathrm{i}\Gamma_{10}$, see Appendix~\ref{app:gamma}. In particular, the symmetric charge conjugation matrix and the block diagonal chirality matrix $\Gamma_*$ are now given by 
\begin{equation}
\Gamma_*=\Gamma_1\Gamma_2\ldots \Gamma_{10}\,,\;\;\;\;\;\;\;\;\mathcal C=\Gamma_{10}\Gamma_*\,,
\end{equation}
Given this real representation for the gamma matrices, it becomes straightforward to identify real sections of the complex Lagrangian \eqref{eq:LSUGRA}. There exists two such sections which are specified by different reality conditions on the spinors and the axions. In both cases the einbein $\mathrm{e}$, the scalars $\phi$ and $V_{i}{}^a$, and the gauge field $A_{ij}$ are real, as they describe the $a_{ijk}=0$ subsector of the one-dimensional theory, which uplifts via \eqref{eq:AnsatzTheta} to the axion-dilaton subsectors of the real Lorentzian type IIB'/IIB$^*$ supergravities.
For the one-dimensional spinor fields and axions, the following choices of reality conditions  
\begin{align}
&1)\;\;\;\psi^*=\psi\,,\;\;\;\;\;\;\;\;\,\;\;\;\lambda^*=\lambda\,,\;\;\;\;\;\;\;\;\;\;\,\;\chi_a^*=\chi_a\,,\;\;\;\;\;a_{ijk}\in \mathbb{R}\,,\nonumber\\[1mm]
&2)\;\;\;\psi^* = -\mathrm{i}\Gamma_*\psi\,,\quad
     \lambda^* = -\mathrm{i}\Gamma_*\lambda\,,\quad
   \chi^a{}^* = \mathrm{i}\Gamma_* \chi^a\,,\;\;\;\;\;a_{ijk}\in \mathrm{i} \mathbb{R}\,,\label{eq:realitycond}
\end{align}
ensure that all the terms in the Lagrangian \eqref{eq:LSUGRA} are real. Note that in the second case the terms in the Lagrangian quadratic in the axions effectively flip sign.

Since we have not worked out the uplift formulas for the axionic sector, it remains \textit{a priori} unclear which of the reality conditions in \eqref{eq:realitycond} correspond to the one-dimensional theories arising from the Kaluza–Klein truncations of type IIB' and type IIB${}^*$ supergravity, respectively. The distinction between these ten-dimensional theories lies in the fact that either the NSNS or the RR two-form is purely imaginary, resulting in an effective sign flip of the associated kinetic term. In Section~\ref{sec:BPSwithaxion}, we analyze supersymmetric solutions of the one-dimensional models characterized by the above reality conditions. For the first choice, we show that the resulting solution makes contact with the so-called spherical brane solutions of type II${}^*$ supergravity \cite{Bobev:2018ugk,Bobev:2024gqg}. We therefore conclude that the first and second choices in \eqref{eq:realitycond} correspond, respectively, to the one-dimensional theories describing truncations of Lorentzian IIB${}^*$ and  IIB' supergravity.

%%%%%%%%%%%%%%%%%%%%%%%%%%%%%%%%%%%%%%%%%%%%%%%

\section{BPS equations, solutions and uplift}\label{sec:BPS}
In this section, we study supersymmetric solutions in the case when $\theta^{ij}$ and $\eta^{ab}$ both have signature (10,0) or (9,1). The analysis is first restricted to the $a_{ijk}=0$ subsector of the one-dimensional theory, and subsequently extended to solutions with a single non-vanishing axion.

\subsection{Supersymmetric solutions without axions}\label{sec:BPSwihtoutaxion}
We consider the supersymmetry transformations of the one-dimensional fields and set the axions $a_{ijk}=0$. In this case, the Killing spinor equations 
\begin{equation}
\delta\lambda\stackrel{!}{=}0\,,
\qquad\delta \chi^a \stackrel{!}{=} 0\,,\label{eq:KS}
\end{equation}
obtained from
(\ref{eq:SUSYall}) with (\ref{eq:YukawaAll}) simply reduce to the following differential equations
\begin{align}
0& =\mathrm{e}^{-1}\dot\phi+\frac{1}{5}\mathrm{g}\,e^{4\phi}\,\mathcal T^{a}{}_a\,,\nonumber\\
0& = \mathrm{e}^{-1}P^{ab}-\mathrm{g}\, e^{4\phi}\big(\mathcal T^{ab}-\frac{1}{10}\eta^{ab}\,\mathcal T^c{}_c\big)\,,\label{eq:BPSa=0}
\end{align}
together with the projection condition on the Killing spinor
\begin{align}
(\mathbb{I}-\Gamma_*\big)\,\epsilon=0\,.\label{eq:projKS}
\end{align}
We have verified that the first-order BPS equations \eqref{eq:BPSa=0} imply all the field equations of the one-dimensional theory. Specifically, taking their derivative leads to the second-order field equations for $\phi$ and the coset scalars $V_i{}^a$, while squaring them yields the first-order equation that results from varying the Lagrangian with respect to the einbein. The field equation for the gauge field $A_{ij}$ follows from dressing the second BPS equation by the coset representative and projecting on the \textbf{45}. The bosonic configurations satisfying \eqref{eq:BPSa=0} therefore correspond to 1/2-BPS solutions of the theory, \textit{i.e.} preserve 16 of the 32 supersymmetries. To satisfy the remaining Killing equation, $\delta\psi\stackrel{!}{=}0$, the 16 Killing spinors defined by \eqref{eq:projKS} must be dressed by a $t$-dependent factor. Note finally that other solutions with vanishing axions would necessarily break all supersymmetries. 

The supersymmetric solutions turn out to uplift to ten-dimensional solutions with flat spacetime geometry and a non-trivial axion-dilaton profile given by harmonic functions. To make this explicit, let us introduce new coordinates $\mathrm{x}_a$, which are related to the embedding coordinates $\mu_i$ satisfying $\theta^{ij}\mu_i\mu_j=1$ and $t$ by
\bea
\mathrm{x}_a=\mathrm{g}^{-1}\,e^{\phi/2}\,V_a{}^i\, \mu_i\,.\label{eq:xa}
%\mu_i&=&\mathrm{g}\,e^{-\phi/2}\,V_i{}^a\,x_a\,.
\eea
This implies in particular,
\bea
\mathrm{g}^2\,\mathcal T^{ab}\,\mathrm{x}_a\mathrm{x}_b=e^\phi\,.
\eea
By using the BPS equations \eqref{eq:BPSa=0}, the relation between covariant differentials can be written as
\begin{equation}
\mathcal D \mathrm{x}_a=\mathrm{g}^{-1}e^{\phi/2}\,V_a{}^iD\mu_i-\mathrm{e}\,e^{9\phi/2}\,\theta^{ij}\,\mu_i V_{ja}\,dt\,,\label{eq:covrelation}
\end{equation}
where the expression for $D\mu_i=D_t\mu_i\,dt$ was given in \eqref{eq:covDTheta}, while
\begin{align}
\mathcal D \mathrm{x}_a&=d\mathrm{x}_a+Q_a{}^{c}\,\mathrm{x}_c\,dt\,,
\end{align}
with the composite $\mathrm{SO}(s,t)_K$ connection $Q^{ab}$ defined in \eqref{eq:PandQ}. Substituting \eqref{eq:covrelation} in the truncation anzatz \eqref{eq:AnsatzTheta} yields the following class of supersymmetric ten-dimensional solutions
\bea
ds_{10}^2&=&\mathcal D \mathrm{x}^a\,\mathcal D \mathrm{x}_a\,,\nonumber\\
e^{-\Phi}&=&\mathrm{g}^2\,e^{3\phi}\,\mathcal T^{ac}\,\mathcal T^{b}{}_c\,\mathrm{x}_{a}\mathrm{x}_b=-\mathcal X\,.\label{eq:upliftBPS1}
\eea
where the one-dimensional fields $\mathcal T^{ab}$ and $\phi$ satisfy \eqref{eq:BPSa=0}.

In order to derive the explicit form of the supersymmetric solutions, we fix the gauge $\mathrm{e}=1$ and pick\footnote{For the theory with (10,0) signature, this is a gauge choice. In the (9,1) signature however, the gauge symmetries are not sufficient to bring $V_i{}^a$ to a diagonal form.} 
\begin{equation}
V_i{}^a=\delta_i^a\,e^{\varphi_a/2}\,,
\end{equation}
as in (\ref{eq:EOMdiag}) above.
Together with $A_{ij}=0$ this implies that $Q^{ab}=0$. Furthermore, upon redefining 
\begin{equation}
    \phi_a := \varphi_a -\phi\,,\;\;\;\;\;\;\;e^\phi=\prod_a e^{-\phi_a/10} \,,
\end{equation}
the first order BPS equations \eqref{eq:BPSa=0} reduce to
\begin{equation}
    \partial_{\tau} {\phi}_a    = -e^{\phi_a}
    \quad\Longrightarrow\quad
    e^{\phi_a}=\frac{1}{\tau+c_a}\,,
    \label{eq:phisol}
\end{equation}
with real constants $c_a$, and where  $\tau(t)$ is a redefined time coordinate that satisfies $\dot \tau=-2\,\mathrm{g}\,e^{5\phi}$.  It is straightforward to verify that this yields a solution to the one-dimensional field equations for any choice of the $c_a$. Let us again emphasize that $\theta^{ij}$ and $\eta^{ab}$ are assumed to both have signature (10,0) or (9,1). The ten-dimensional uplifts \eqref{eq:upliftBPS1} then read
\bea
ds_{10}^2&=&d \mathrm{x}^a\,d \mathrm{x}_a\,,\nonumber\\[1ex]
e^{-\Phi}&=&\mathrm{g}^2\,e^{5\phi}\,\sum_a e^{2\phi_a}\,\mathrm{x}^{a}\mathrm{x}_a=-{\cal X}\,.\label{eq:generalsup}
\eea
with the explicit form of the functions $\phi_a$ in \eqref{eq:phisol}. The metric describes $\mathbb{R}^{s,t}$ and the ten-dimensional dilaton $e^\Phi$ satisfies Laplace's equation.
When all the constants $c_a$ are equal, the solution reduces to \eqref{eq:solutionSignature}, though expressed in a different gauge for the one-dimensional diffeomorphisms. For the (10,0) signature, it corresponds to the (near-horizon geometry) of the D$(-1)$ brane solution. More generally in this signature, for $I$ sets of $n_I\geq 2$ equal constants, we expect these solutions to be the analogues of the smeared D3-brane solutions studied in \cite{Freedman:1999gk}, describing $\prod_I {\rm SO}(n_I)$-invariant distributions of $N$ instantons spread across the nine transverse directions. The precise distributions of the instantons are encoded in the expression of the dilaton $e^{\Phi}$.

\subsection{Supersymmetric solutions with axions}\label{sec:BPSwithaxion}

Let us now consider supersymmetric solutions in the presence of non-vanishing axion fields. A non-vanishing axion field $a_{ijk}$ necessarily breaks the gauge symmetry. In a minimal setting, we may switch on a single axion field, say, $y=a_{8,9,10}$, thus breaking the gauge group down to 
\bea
    {\rm SO}(10)_\mathrm{g} &\longrightarrow&  {\rm SO}(7)\times {\rm SO}(3) \,,
    \nonumber\\
    {\rm SO}(9,1)_\mathrm{g} &\longrightarrow& {\rm SO}(7) \times {\rm SO}(2,1)\,,
    \label{eq:SO37}
\eea
for the two cases when $\theta^{ij}$ and $\eta^{ab}$ both have signature (10,0) or (9,1). 
The smaller symmetry groups also allow for a non-trivial scalar matrix $T_{ij}$ of the form
\begin{equation}
    T_{ij} = {\rm diag}\big( e^{-3x}, e^{-3x}, e^{-3x}, e^{-3x}, e^{-3x}, e^{-3x}, e^{-3x},  e^{7x}, e^{7x}, \pm e^{7x}\big)\,.
\end{equation}
On the other hand, all vector fields are set to zero. In the rest of this discussion we restrict to the case of (9,1) signature, i.e.\ to the theories descending from IIB$^*$ and IIB' supergravity, c.f.\ Section \ref{sec:realforms}.
In this truncation, the Lagrangian (\ref{eq:LSUGRA}) reduces to
\bea
\mathcal{L}&=&10\,\mathrm{e}^{-1}\, \dot \phi^2
-\frac{105}{2}\,\mathrm{e}^{-1}\,\dot x^2
+\frac1{2}\,\mathrm{e}^{-1}\,e^{-2\phi-21x} \,\dot y^2
\nonumber\\
&&{}
+\frac12\, \mathrm{e}\,\mathrm{g}^2 \,e^{8\phi}\left(35\,e^{-6x}+42\,e^{4x}+3\,e^{14\,x}\right)
-\frac32\,\mathrm{e}\,\mathrm{g}^2\,e^{6\phi-7x}\,y^2
\,,
\label{eq:L37ax}
\eea
for the remaining fields $\{\mathrm{e}, \phi, x, y\}$. The Killing spinor equations \begin{equation}
\delta\lambda\stackrel{!}{=}0\,,
\qquad\delta \chi^a \stackrel{!}{=} 0\,,
\end{equation}
obtained from \ref{eq:SUSYall}) with (\ref{eq:YukawaAll}) 
yield the differential equations
\bea
0 &=&   
   100\, e^{7 x}
    \left(e^{x+2\phi} \,\big(e^{10 x}+3\big)^2 -y^2\right) \dot x^2
-
\mathrm{g}^2 e^{6 \phi} \left(2\,e^{x+2\phi}\, \left(e^{10 x}-1\right)
   \left(e^{10 x}+3\right) +y^2 \right)^2
   ,
 \nonumber\\
 0 &=& 20 \,y  \left(3 e^{10 x}+1\right) e^{11 x+2 \phi} \,\dot\phi- \left(9\, y^2-4\,
   \big(e^{10 x}+3\big)\, \big(3 e^{10 x}+7\big)\, e^{x+2 \phi }\right)\dot y
   \,,
\nonumber\\
0 &=& 20 \,y  \left(3 e^{10 x}+1\right) e^{11 x+2 \phi } \,\dot x- \left(2\, y^2+4 \,
\big(e^{10 x}-1\big)
   \big(e^{10 x}+3\big) e^{x+2 \phi}\right)\dot y
   \,,
  \label{eq:BPS-BS}
\eea
together with the following projection condition on the Killing spinor
\begin{equation}
 \mathbb{P}\,\epsilon=0\,,\qquad   \mathbb{P}
    \equiv \frac12\,(\mathbb{I}-\Gamma_*)
    \Big(\mathbb{I}+
    \frac{e^{-\phi-21 x/2} \,\big(9\, \mathrm{g}\,
    e^{7 x+4 \phi}\,y+\dot y\big)}{4 \,\big(\mathrm{g}\, \big(3 \,e^{7
   x}+7\,e^{-3x}\big)\, e^{4 \phi}-5\,  \,\dot\phi\big)}\,
    \widehat\Gamma\Big)
\,,
\label{eq:projector}
\end{equation}
with $\widehat\Gamma=\Gamma_8\Gamma_9\Gamma_{10}$.
It is straightforward to check that $\mathbb{P}$ is indeed a projector
\begin{equation}
\mathbb{P}^2=\mathbb{P}\,,\qquad
{\rm tr}\,\mathbb{P} = 16\,,
\end{equation}
showing that solutions of (\ref{eq:BPS-BS}) preserve half of all supersymmetries.\footnote{The journal version of Ref.\ \cite{Ciceri:2025maa} over-restrictively stated that 1/2-BPS solutions with non-vanishing axions were excluded.} Finally, one can show that equations (\ref{eq:BPS-BS}) also imply that
\begin{equation}
    \delta_\epsilon \psi= 0\,,
\end{equation}
upon dressing the Killing spinors defined by (\ref{eq:projector}) with a further appropriate $t$-dependent factor.
In order to analyze the space of solutions of (\ref{eq:BPS-BS}), it is useful to relate our equations to the analysis of \cite{Bobev:2018ugk,Bobev:2024gqg}. To this end, we define the fields
\begin{equation}
    X=e^{10x}\,,\quad Y=e^{\phi+x/2}\,y\,,\quad
    \eta=\frac72\,(x+2\,\phi)
    \,,
    \label{eq:XYeta}
\end{equation}
in terms of which the Lagrangian (\ref{eq:L37ax}) turns into
\bea
    {\cal L} &=& -\frac12\,\mathrm{e}^{-1}\,\frac{\dot{X}^2-\dot{Y}^2}{X^2}
    -\frac17\,\mathrm{e}^{-1}\,\frac{X\dot{X}-Y\dot{Y}}{X^2}\,\dot{\eta}
    +\frac{10}{49}\,\mathrm{e}^{-1}\,(1+Y^2X^{-2})\,\dot\eta^2
    \nonumber\\
    &&{}
    + \frac12\,\mathrm{g}^2\,\mathrm{e}\,e^{8\eta/7}\,
    \frac{35+42\,X+3\,X^2-3\,Y^2}{X}
    \,.
\eea
This is precisely the Lagrangian of \cite{Bobev:2024gqg}, equation (B.1), extrapolated to $d=0$\,.
Its equations of motion are implied by the first order BPS equations given in (B.8) in \cite{Bobev:2024gqg}
\bea
\frac{dY}{dX} &=& - \frac{Y \left(7 X^2-Y^2+8 X+9\right)}{2 X \left(2 (1-X) (X+3)-Y^2\right)}\,,
   \nonumber\\
\frac{d\eta}{dX} &=& \frac{7 \left((X+3)^2-Y^2\right)}{2 X \left(2 (1-X) (X+3)-Y^2\right)}
\,,
\label{eq:BPS-B}
\eea
upon setting ${\rm e}=1$, together with
\begin{equation}
    (3+X-Y) (3+X+Y) \,\dot{X}^2
   =
   \mathrm{g}^2 e^{\frac{8 \eta}{7}} X
   \left(2 X
   (X+2)+Y^2-6\right)^2
   \,,
   \label{eq:BPS-B2}
\end{equation}
to recover the original coordinate $t$. With (\ref{eq:XYeta}), these equations
indeed coincide with the Killing spinor equations (\ref{eq:BPS-BS}) above. We thus find perfect agreement with the equations of \cite{Bobev:2024gqg}.

Let us stress that the BPS equations (\ref{eq:BPS-B}), (\ref{eq:BPS-B2}) admit solutions with real or purely imaginary $Y$ field. The former correspond to solutions of the Lagrangian (\ref{eq:L37ax}) with a real axion $y=a_{8,9,10}$, and reproduce the configurations analyzed in \cite{Bobev:2024gqg}, which uplift to the $p=-1$ analytic continuation of the spherical D$p$-brane solutions of type IIB${}^*$ supergravity. Instead, the theory described by the Lagrangian (\ref{eq:L37ax}) with an imaginary axion $y$ descends from the truncation of type IIB' supergravity. This establishes the lower-dimensional identification of Lorentzian IIB' versus IIB${}^*$ supergravity in terms of the reality properties of the axions $a_{ijk}$, cf.\ Section~\ref{sec:realforms}.

Finally, the above analysis can be repeated for the one-dimensional theory with (10,0) signature, descending from Euclidean type IIB supergravity. The theory also admits supersymmetric solutions defined by (\ref{eq:BPS-B}), (\ref{eq:BPS-B2}).

%%%%%%%%%%%%%%%%%%%%%%%%%%%%%%%%%%%%%%%%%%%%%%%

\section{Conclusions and outlook}\label{sec:Conclusion}

In this paper we have presented the detailed construction of maximal ${\rm SO}(10)_\mathrm{g}$ gauged supergravity in one dimension which is expected to arise from consistent truncation of Euclidean IIB supergravity on a sphere $S^9$. We have provided the explicit reduction ans\"atze for the bosonic ${\rm SL}(10)/{\rm SO}(10)_K$ subsector of the one-dimensional theory which uplifts to the ten-dimensional dilaton-axion system. Starting from this subsector, we have constructed the full one-dimensional theory by imposing maximal supersymmetry. The field content of this theory precisely matches the lowest BPS multiplet of single trace operators from \eqref{eq:O123}, as
\begin{align}
{\cal T}_{ab}
\; \longleftrightarrow\;\,& {\cal O}^{ab} = {\rm Tr}[X^{a} X^{b}]-\tfrac1{10}\,\delta^{ab}\,{\rm Tr}[X^{c} X^{c}] 
\,,\nonumber\\
\chi_a\; \longleftrightarrow\;\,&{\cal O}^a = {\rm Tr}[X^a\,\Psi] - \tfrac19\, {\rm Tr}[X_b\,\Gamma^{ab} \Psi]
\,,\nonumber\\
a_{abc}\; \longleftrightarrow\;\,&{\cal O}^{abc} = {\rm Tr}\left[X^{a} [X^{b},X^c]\right] -\tfrac18\, {\rm Tr}\left[\bar\Psi \Gamma^{abc} \Psi \right]
\,,
\end{align}
for the propagating fields. The one-dimensional theory provides the starting point for setting up the procedure of holographic renormalization following \cite{Bianchi:2001kw,Wiseman:2008qa,Kanitscheider:2008kd}, thereby preparing the ground for the holographic computation of correlation functions in the IKKT model.
It would be most interesting to compare these results with numerical studies of the model. 

An open question concerns the existence of an analogous framework for the holographic description of the Lorentzian IKKT model. The dual background corresponds to the flat $\mathrm{SO}(9,1)$-invariant solution discussed in \cite{Blair:2025nno}, which sits in the hyperboloid truncation of IIB${}^*$ supergravity, see \eqref{eq:solutionSignature}. In particular, such a framework would have to capture the qualitative differences of the Lorentzian matrix model, revealed in recent numerical simulations \cite{Hatakeyama:2021ake,Asano:2024def,Chou:2025moy}.

By itself, the one-dimensional maximal supergravity stands out as the first of its kind, while supergravity theories in all higher dimensions $d$ have been systematically explored\footnote{With the exception of $d=2$, where only those maximal supergravities arising from consistent Kaluza–Klein truncations of IIA/IIB supergravity have been constructed \cite{Bossard:2023wgg}.}. A systematic analysis of one-dimensional maximal supergravities remains a challenging task. In particular, the ${\rm SO}(10)_\mathrm{g}$ gauged model has been constructed directly from its subsector (\ref{eq:L1Eu}) obtained from reduction on the sphere $S^9$. The limit of vanishing coupling constant $\mathrm{g}\rightarrow0$ presents a number of subtleties. Although the Lagrangian (\ref{eq:LSUGRA}) at first view appears to admit a smooth limit $\mathrm{g}\rightarrow0$, in this limit the constraints arising as the gauge field equations disappear from the theory, thus altering the number of propagating degrees of freedom. The resulting theory no longer admits an immediate higher-dimensional origin. This is quite different from the situation in higher dimensions, where the ungauged limit of all supergravities recovers the toroidally reduced theory with its characteristic symmetry enhancement to the full exceptional symmetry group ${\rm E}_{11-d,(11-d)}$. If this pattern were to persist down to one dimension, the resulting theory should provide a concrete realization of the hyperbolic group ${\rm E}_{10(10)}$ in accordance with the longstanding conjecture of \cite{Julia:1982gx}, and related constructions in \cite{Mizoguchi:1997si,Damour:2002cu}.

An embedding of our construction into the structures of exceptional geometry should also streamline the derivation of the explicit uplift formulae to higher dimensions. Here, we have given explicit uplift formulae for the ${\rm SL}(10)/{\rm SO}(10)_K$ subsector, but for non-vanishing axions $a_{ijk}$ we expect highly non-linear reduction ansätze for the ten-dimensional p-forms. A derivation of these formulae would require the embedding of the reduction into some exceptional field theory, as has been realized in other dimensions~\cite{Baguet:2015sma,Bossard:2023jid}. The master formulation  \cite{Bossard:2021ebg} may provide valuable guidance for this endeavour. In the longer term, such a construction may ultimately reveal the imprints of the exceptional symmetry in the spectrum of the matrix model along the lines established in higher dimensions \cite{Malek:2019eaz}.

Through the analysis of the one-dimensional Killing spinor equations, we have identified two main classes of 1/2-BPS solutions. The first uplifts to a class of IIB solutions with flat spacetime metric and axion/dilaton profiles governed by harmonic functions, while all other p-forms vanish. This class contains the near-horizon limit of the D$(-1)$-instanton as a distinguished $\mathrm{SO}(10)$-invariant solution. More generally, we expect these solutions to describe continuous distributions of D-instantons along the nine transverse directions, thereby generically breaking $\mathrm{SO}(10)$, in direct analogy with the patterns observed in higher-dimensional D$p$-brane holographic setups, see for instance \cite{Freedman:1999gk}. The second class of 1/2-BPS solutions in the one-dimensional theory involves non-vanishing axions $a_{ijk}$, and is expected to uplift to IIB configurations with non-trivial profiles for the p-form gauge fields. For the real forms of the theory in (9,1) signature, these solutions preserve an $\mathrm{SO}(7)\times \mathrm{SO}(2,1)$ isometry. We have shown that a subset of them corresponds to (an analytic continuation of) the spherical brane solutions analyzed in \cite{Bobev:2018ugk,Bobev:2024gqg}, which are known to holographically describe supersymmetric mass deformations of the dual boundary theory. It would be interesting to explore in more detail the analogous solutions of the theory in (10,0) signature, and to investigate potential connections to the supersymmetric backgrounds of Euclidean IIB supergravity constructed in \cite{Komatsu:2024bop}, which are dual to vacua of the polarized IKKT matrix model \cite{Bonelli:2002mb} and have recently served as the basis for concrete tests of ‘timeless holography’ \cite{Hartnoll:2024csr,Komatsu:2024ydh,Komatsu:2024bop}. It would also be worthwhile to examine potential relations to the geometries of \cite{Lin:2005nh}. Another interesting direction would be to explore connections to the holographic setup based on the D$(-1)$/D7 system \cite{Billo:2021xzh}, whose near-horizon background exhibits similarities with some of our solutions \cite{Aguilar-Gutierrez:2022kvk}.

We hope to come back to some of these issues in the future.

\subsection*{Acknowledgements}
We would like to thank Nikolay Bobev, Guillaume Bossard, Fran\c{c}ois Delduc, Axel Kleinschmidt, and Adrien Martina for useful discussions. The research of FC is funded by the Deutsche Forschungsgemeinschaft (DFG, German Research Foundation) – Project number: 521509185.

\section*{Appendix}

\begin{appendix}

\section{$\Gamma$-matrices conventions}\label{app:gamma}
%%%%%%%%%%%%%%%%%%%%%%%%
We start by considering $32\times 32$ complex gamma matrices $\Gamma_a$ satisfying the standard $\mathrm{SO}(10)$ Clifford algebra
\bea
\{\Gamma_a,\Gamma_b\}=2\,\delta_{ab}\,\mathbb{I}_{32}\,.
\eea
It is useful to have an explicit representation of these gamma matrices. Following the Appendix A of \cite{Komatsu:2024ydh}, we choose a Weyl basis in which
\bea
\Gamma_{a}:=\begin{pmatrix} 0 &\gamma_a \\ \bar\gamma_a & 0 \end{pmatrix}\,,\label{eq:GammaM}
\eea
and where $\gamma_{a}$ are $16\times 16$ symmetric matrices. The bar denotes complex conjugation. In particular, we have 
\bea
\gamma_{10}:=-\mathrm{i}\mathbb{I}_{16}
\eea
while the $\mathrm{SO}(9)$ matrices $\gamma_p$, with $p=1,\dots,9$ are real and satisfy
\bea
\{\gamma_p,\gamma_q\}=2\,\delta_{pq}\,\mathbb{I}_{16}\,. 
\eea
We can pick
\bea
\gamma_1:=&\sigma^{2}\otimes\sigma^{2}\otimes\sigma^{1}\otimes\mathbb I\,,\nonumber\\
\gamma_2:=&\sigma^{2}\otimes\sigma^{3}\otimes\mathbb I\otimes \sigma^{2}\,,\nonumber\\
\gamma_3:=&-\sigma^{3}\otimes\mathbb I\otimes \mathbf 1\otimes \mathbb I 1\,,\nonumber\\
\gamma_4:=&-\sigma^{2}\otimes \sigma^{1}\otimes \mathbb I\otimes \sigma^{2}\,,\nonumber\\
\gamma_5:=&\sigma^{2}\otimes \mathbb I\otimes \sigma^{2}\otimes \sigma^{3}\,,\nonumber\\
\gamma_6:=&\sigma^{1}\otimes \mathbb I\otimes \mathbf 1\otimes \mathbb I\,,\nonumber\\
\gamma_7:=&\sigma^{2}\otimes \sigma^{2}\otimes \sigma^{2}\otimes \sigma^{2}\,,\nonumber\\
\gamma_8:=&-\sigma^{2}\otimes \mathbb I\otimes \sigma^{2}\otimes \sigma^{1}\,,\nonumber\\
\gamma_9:=&\sigma^{2}\otimes \sigma^{2}\otimes \sigma^{3}\otimes \mathbb I\,,
\eea
where the $\sigma$'s denote Pauli matrices. Overall, this means that we have 
\bea
\Gamma_a^\dagger=\Gamma_a\,,\;\;\;\;\;\;\gamma_p^\dagger=\gamma_p\,.
\eea
In this Weyl basis, the chirality gamma matrix $\Gamma_{*}$ is block diagonal,
\bea
\Gamma_{*}=-\mathrm{i}\Gamma_{1}\ldots\Gamma_{10}=\begin{pmatrix} \mathbb I_{16} &0 \\ 0 & -\mathbb I_{16}\label{eq:Gamma*} \end{pmatrix}\,.
\eea
It anticommutes with all gamma matrices and satisfies $\Gamma_{*}^2=\mathbb{I}_{32}$. Finally we choose a symmetric\footnote{This differs from \cite{Komatsu:2024ydh}, where the charge conjugation is antisymmetric and imaginary $\mathcal C_{\text{there}}=\mathrm{i}\,\mathcal C_{\text{here}}\,\Gamma_*$.} charge conjugation matrix,
\bea
\mathcal C=-\mathrm{i}\Gamma_{10}\Gamma_{*}=\begin{pmatrix}0 &\mathbb I_{16} \\ \mathbb I_{16} &0\end{pmatrix}\,,
\label{eq:CC}
\eea
which satisfies
\bea
\mathcal C=\bar{\mathcal C}=\mathcal C^T=\mathcal C^{-1}\,,\;\;\;\;\;\mathcal C^2=\mathbb{I}_{32}\,,\;\;\;\;\;\Gamma_a^T=\mathcal C\Gamma_a\mathcal C^{-1}\,.\label{eq:Csymm}
\eea
The last equation implies the following symmetry properties for antisymmetric products of gamma matrices contracted with $\mathcal C$,
\bea
\text{symmetric}&&:\;\mathcal C\,,\mathcal C\Gamma_a\,,\mathcal C\Gamma_a\Gamma_*\,,\mathcal C\Gamma_{ab}\Gamma_{*}\,,\mathcal C \Gamma_{abcd}\,,\mathcal C \Gamma_{abcde}\,,\nonumber\\
\text{antisymmetric}&&:\; \mathcal C\Gamma_{*}\,,\mathcal C\Gamma_{ab}\,,\mathcal C\Gamma_{abc}\,,\mathcal C\Gamma_{abc}\Gamma_{*}\,,\mathcal C\Gamma_{abcd}\Gamma_*\,.\label{eq:Gammasym}
\eea

In Sections~\ref{sec:SUGRA} and~\ref{sec:BPS}, which deal with one-dimensional supergravities, we work with 32-component spinors $\psi^\alpha$ that decompose as
\bea
\psi^\alpha=\begin{pmatrix}\psi^I\\ \psi_I\end{pmatrix}\,,
\eea 
with $\alpha\,,\beta=1,\ldots,32$ and $I,J=1,\ldots,16$. We also define the charge conjugated spinor as
\bea
 \bar\psi_\alpha:=\psi^\beta\,\mathcal C_{\beta\alpha}\,,\;\;\;\;\;\text{such that\;\;}\;\;\;\bar\psi_\alpha=\big(\psi_I\;\;\;\psi^I\big)\,.
 \eea 
 The spinor components are anticommuting Grassmann numbers. Together with the symmetry properties \eqref{eq:Gammasym}, this leads to identities of the form 
 \bea
\bar\psi\,\lambda=-\bar\lambda\,\psi\,,\;\;\;\;\;\;\bar\psi\,\Gamma_*\,\lambda=\bar\lambda\,\Gamma_*\,\psi\,,\;\;\;\;\;\;\bar\psi\,\Gamma_{abc}\lambda=\bar\lambda\,\Gamma_{abc}\psi\,,\;\;\;\;\;\;\ldots
 \eea
 for the fermions bilinears. In contrast to the supergravity spinors, the $\mathrm{SO}(10)$ spinor matrix of the IKKT model in Section~\ref{sec:IKKT} satisfies the chirality constraint $\Gamma_*\Psi=\Psi$. This implies
\bea
\Psi^\alpha=\begin{pmatrix}\Psi^I\\ 0\end{pmatrix}\,, \;\;\;\;\;\;\;\bar\Psi_\alpha=\big(0\;\;\;\Psi^I)\,.
\eea
With explicit indices, the antisymmetric products of $\mathrm{SO}(10)$ gamma matrices and the charge conjugation read
\bea
(\Gamma_{*})^\alpha{}_\beta &=&\begin{pmatrix}\delta^I{}_J &0 \\ 0& -\delta_I{}^J \end{pmatrix}\,,\;\;\;\mathcal C_{\alpha\beta}=\begin{pmatrix} 0 &\delta^{IJ} \\ \delta_{IJ} & 0 \end{pmatrix} \,,\nonumber\\[2mm]
(\Gamma_{a})^\alpha{}_\beta &=&\begin{pmatrix} 0 &\gamma_a{}^{(IJ)} \\ \bar\gamma_{a\,(IJ)} & 0 \end{pmatrix}\,,\;\;\;(\Gamma_{ab})^\alpha{}_\beta=\begin{pmatrix} \gamma_{ab}{}^I{}_J &0 \\ 0& \bar\gamma_{ab\,I}{}^J  \end{pmatrix}\,,\;\;\;\ldots
\eea
where $\gamma_{ab}:=\gamma_{[a}\bar\gamma_{b]}$ satisfies $\gamma_{ab}^T=-\bar\gamma_{ab}$. 
Note also that under an ${\rm SO}(10)$ transformation with parameter $\omega^{ab}$, we have 
\bea
\delta_\omega \psi^\alpha=\omega^{ab}\,(\Gamma_{ab})^{\alpha}{}_{\beta}\psi^\beta\,,\;\;\;\text{and}\;\;\;\;\delta_\omega \bar\psi_\alpha=-\omega^{ab}\,\bar\psi_\beta\,(\Gamma_{ab})^{\beta}{}_{\alpha}\,,
\eea
such that the following bilinears are obviously invariant
\begin{align}
\bar\psi\,\lambda =&\,\psi^I\lambda_I+\psi_I\lambda^I \,,\nonumber\\
\bar\psi\,\Gamma_{*}\,\lambda =&-\psi^I\lambda_I+\psi_I\lambda^I \,.
\end{align}

Let us also discuss our conventions for SO(9,1) gamma matrices. They satisfy
\bea
    \{\Gamma_a,\Gamma_b\}=2\,\eta_{ab}\,\mathbb{I}_{32}\,,
\eea
with $\eta_{ab}=\text{diag}(+,+,\ldots, -)$. An explicit Weyl representation is obtained from the above SO(10) gamma matrices by sending
\bea
    \Gamma_{10}\longrightarrow -\mathrm i \Gamma_{10}=\begin{pmatrix}0 & -\mathbb{I}_{16}\\
    \mathbb{I}_{16} & 0\end{pmatrix}\,.
\eea
such that now $\Gamma_{10}^\dagger=-\Gamma_{10}$. Note that in this Weyl representation all the matrices are real. The expression of the charge conjugation matrix \eqref{eq:CC} and the chirality matrix \eqref{eq:Gamma*} remain the same, and are now given in terms of the $\mathrm{SO}(9,1)$ gamma matrices by 
\bea
\mathcal C=\Gamma_{10}\Gamma_*\,,\;\;\;\;\;\Gamma_*=\Gamma_1\ldots\Gamma_{10}\,.
\eea
The symmetry properties \eqref{eq:Csymm} and \eqref{eq:Gammasym} still hold for the $\mathrm{SO}(9,1)$ matrices.

Finally, the duality relations between the above $\mathrm{SO}(s,t)$ gamma matrices, both for $s=10,t=0$ and $s=9,t=1$, can be summarized by 
\begin{equation}
\Gamma_{a_1\ldots a_p}\Gamma_*=\tfrac{\mathrm{i}^{(s+1)}(-1)^{[p/2]}}{(10-p)!}\,\varepsilon_{a_1\ldots a_pa_{p+1}\ldots a_{10}}\,\Gamma^{a_{p+1}\ldots a_{10}}\,,\label{eq:gammadual}
\end{equation}
where $[\cdot]$ takes the integer part, and with the $\mathrm{SO}(s,t)$ Levi-Civita tensor $\varepsilon_{1\ldots 10}=1$. 
%$\varepsilon^{1\ldots 10}=1$.
\section{${\rm SO}(10)$ representations}
\label{app:SO10}

In the main text, we specify ${\rm SO}(10)$ representations by their Dynkin labels
\begin{equation}
    [n_1 n_2 n_3 n_4 n_5]\,,\qquad n_i \in \mathbb{N}
    \,.
\end{equation}
For the representations of lowest dimensions, we will equivalently denote them by their dimension
\bea
[00000] &:& {\bf 1}
\,,\nonumber\\{}
[10000] &:& {\bf 10}
\,,\nonumber\\{}
[20000] &:& {\bf 54}
\,,\nonumber\\{}
[01000] &:& {\bf 45}
\,,\nonumber\\{}
[00100] &:& {\bf 120}
\,,\nonumber\\{}
[00001] &:& {\bf 16}_s
\,,\nonumber\\{}
[00010] &:& {\bf 16}_c
\,,\nonumber\\{}
[10010] &:& {\bf 144}_c
\,.
\eea

The generic representations that appear in the tower of BPS multiplets (\ref{eq:towerBPS}) have dimensions
\bea
{}
[n,0000] &:&
\frac1{20160}\,
(n+1) (n+2) (n+3) (n+4)^2 (n+5) (n+6) (n+7)
\,,\nonumber\\{}
[n,0001] &:&
\frac1{2520}\,
(n+1) (n+2) (n+3) (n+4) (n+5) (n+6) (n+7) (n+8)
\,,\nonumber\\{}
[n,0100] &:&
\frac{1}{720}\, (n+1) (n+2) (n+4) (n+5)^2 (n+6) (n+8) (n+9)
\,,\nonumber\\{}
[n,1010] &:&
\frac{1}{360}\, (n+1) (n+3) (n+4) (n+5) (n+6) (n+7) (n+8) (n+10)
\,,\nonumber\\{}
[n,2000] &:&
\frac{1}{576}\, (n+1) (n+4) (n+5) (n+6)^2 (n+7) (n+8) (n+11)
\,,\nonumber\\{}
[n,0020] &:&
\frac{1}{576}\, (n+1) (n+2) (n+3) (n+4) (n+6) (n+7) (n+8) (n+9)
\,.
\label{eq:SO10dims}
\eea
Formally, these dimension formulas can be extended to negative Dynkin weights and suggest the identifications
\bea
[-3, 2 0 0 0] &:& -{\bf 10}
\,,\nonumber\\{}
[-2, 2 0 0 0] &:& -{\bf 45}
\,,\nonumber\\{}
[-2, 1 0 1 0]  &:& -{\bf 16}_c
\,,\nonumber\\{}
[-3, 0 1 0 0]  &:& {\bf 1}
\,.
\label{eq:negn}
\eea
The identification (\ref{eq:negn}) can in fact be taken beyond mere numerology and extends to the level of the full character polynomials. In the main text, we will see that these identifications capture the concrete physical realization of the lowest BPS multiplets.

The character polynomials for the lowest ${\rm SO}(10)$ representations are given by
\bea
\chi_{\bf 10} &=& \sum_{i=1}^5\big(y_i+y_i^{-1}\big)\,,
\nonumber\\
\chi_{{\bf 16}_s} &=&
\sum_{\alpha=1}^{16} {\bf y}^{\bf q_{\alpha}} \,,
\nonumber\\
\chi_{{\bf 16}_c} &=&
\sum_{\alpha=1}^{16} {\bf y}^{-\bf q_{\alpha}} \,,
\eea
with
\bea
{\bf y}^{\bf q_{\alpha}}=\prod_{i=1}^5\,y_{i}^{q_{\alpha i}}\,,\qquad
{\bf q_{\alpha}}\in
\Big\{ \big\{\pm\tfrac12,\pm\ft12, \pm\ft12,\pm\ft12,\pm\ft12\big\}  
\;\Big|\;  \; {\rm with~even~number~of~+} \Big\}
\;,
\eea
in terms of the five charges $y_i$. The full Polya formula for counting cyclic words is then given by
\begin{equation}
{\cal Z}_{\rm IKKT}(t, y_i)
=
  -\sum_{m=1}^\infty \frac{\varphi(m)}{m}\,\log\left[1-{\bf Z_1}(t^m, y_i^m) \right]
  \;,
\label{eq:ZIKKT-CP}
\end{equation}
with the full single letter partition function
\begin{equation}
{\bf Z_1}(t, y_i)\equiv \chi_{\bf 10}\, t -\chi_{{\bf 16}_s}\, t^{3/2} +\chi_{{\bf 16}_c}\,
t^{5/2}-\chi_{\bf 10}\, t^3+ t^4
\;.
\label{eq:z1CP} 
\end{equation}
Expansion of (\ref{eq:ZIKKT-CP}) allows to determine the full ${\rm SO}(10)$ representation content of the spectrum of single trace operators.

\section{Linear and quadratic relations among Yukawa tensors}\label{app:Yukawa}

We start by collecting the linear relations among the Yukawa tensors $A, B, C^a, E, E^a, E^{ab}$ that follow from requiring the supersymmetry variations of the Lagrangian \eqref{eq:LSUGRA} to vanish at order $\mathrm{g}$. These variations always contain a single derivative acting on a bosonic field and a fermion bilinear involving the supersymmetry parameter together with one of the fermions
$\psi,\lambda$ or $\chi_a$. 

The variations proportional to $\dot \phi$ lead to the relations
\begin{flalign}
40\frac{\partial  B}{\partial \phi}-20\,\Gamma_*\, A+ E\,\Gamma_*&=0\,,\nonumber\\%\;\;\;\;\;\;\;\;{\lambda}\\
(\mathcal C\,\Gamma_* B)_{(\alpha\beta)}&=0\,,\nonumber\\%\;\;\;\;\;\;\;\;{\psi}\\
8(\mathcal C \frac{\partial C^a}{\partial \phi})_{\alpha\beta}+(\mathcal C\Gamma_* E^a)_{\beta\alpha}&=0\,,%\;\;\;\;\;\;\;\;\chi_{a}^{\alpha}
\end{flalign}
where in the last equation one must bear in mind the trace condition \eqref{eq:tracechi} on the indices ${}^{a}_{\alpha}$ coming from the implicit contraction with $\chi_{a}^{\alpha}$. For the variations proportional to the current $P_{ab}$, we get
\begin{flalign}
\big(\mathcal C\,\Gamma^{(a} C^{b)}\big)_{(\alpha\beta)}+\frac{1}{2}e^{4\phi}\mathcal T^{c(a}\left((\mathcal C\,\Gamma^{b)}{}_{c}\Gamma_*)_{\alpha\beta}+e^{-\phi}a^{b)}{}_{cd}(\mathcal C\Gamma^d\Gamma_*)_{\alpha\beta}\right)&=0\,,\nonumber\\[2mm]%\;\;\;\;\;\;\;\psi[4mm]
 E^{(a}\Gamma^{b)}+80\left(\frac{\partial B}{\partial \Sigma^{ab}}+3a^{cd(a}\frac{\partial B}{\partial a_{b)}{}^{cd}}\right)-16\,e^{4\phi}\mathcal T^{c(a}\left(\Gamma^{b)}{}_{c}-\frac{3}{4}e^{-\phi}\,a^{b)}{}_{cd}\,\Gamma^d\right)&=0\nonumber\\[2mm]%\,,\;\;\;\;\;\;\;\lambda\\[4mm]
\mathcal P_{bc}\,\Gamma^cE^{ba}+2\,\mathcal P^{ab}\,\Gamma_b A-4\,\mathcal P_{bc}\left(\frac{\partial  C^a}{\partial\Sigma^{bc}}+3\,a^{efb}\frac{\partial C^a}{\partial a^{cef}}\right)&\nonumber\\[1mm]
+4\,e^{4\phi}\mathcal P^{c[b}\,\mathcal T^{a]}{}_{c}\,\Gamma_b\Gamma_*+2\,e^{3\phi}\,\mathcal P_{bc}\,\mathcal T^{dc}\,a^{ab}{}_{d}\,\Gamma_*&=0\,.\label{eq:linPpsi}%\;\;\;\;\;\;\;\chi_{a}^{\alpha}
\end{flalign}
Note that in the first two equations the indices $(ab)$ are projected onto the $\mathbf{54}$ by the current. In the last equation, which is implicitly acted upon by $\bar\chi_a$ from the left, we kept the current explicit for clarity. The terms that do not involve Yukawa tensors may be viewed as ‘source terms’, arising from the variation of the gauge fields inside the $\sim P_{ab}P^{ab}$ kinetic term of the Lagrangian. We also introduced the $\mathrm{SO}(s,t)_K$ covariant variation $\partial/\partial\Sigma^{ab}$ with respect to the 54 coset scalars, where $\Sigma^{ab}$ is defined by
\begin{equation}
\delta V_i{}^a=:\Sigma^{ab}\,V_{ib}\,.
\end{equation}
Finally, for the variations proportional to the current $p_{abc}$ we find
 \begin{flalign}
2\,(\mathcal C\Gamma^{bc} C^a)_{(\alpha\beta)}+\frac{2}{3}\,(\mathcal C\Gamma^{abc}\Gamma_* B)_{(\alpha\beta)}
-e^{3\phi}\,\mathcal T_{de}\,a^{eab}\left((\mathcal C\Gamma^{cd}\Gamma_*)_{\alpha\beta}+e^{-\phi}a^{cdf}(\mathcal C\Gamma_{f}\Gamma_*)_{\alpha\beta}\right)&=0\,,\nonumber\\[4mm]%\,,\;\;\;\;\psi\nonumber\\[4mm]
80\,e^\phi\frac{\partial B}{\partial a_{abc}}-\frac13\Gamma^{abc}\Gamma_* A-4\Gamma^{abc} B+2\,\Gamma^{ab}\Gamma_* C^c-\frac{10}{3} B\Gamma^{abc}+\frac{1}{60} D\Gamma^{abc}\Gamma_*&\nonumber\\
+\frac{7}{40}E_d\tilde\Gamma^{da,bc}-4\,e^{3\phi}\mathcal T_{de}\,a^{eab}\left(\Gamma^{cd}+\frac{3}{4}e^{-\phi}\,a^{fdc}\,\Gamma_f\right)+\frac{1}{24} b_{defg}\,\Gamma^{abcdefg}&=0\,,\nonumber\\[4mm]%\,,\;\;\;\;\lambda\nonumber\\[4mm]
\left(16\,e^\phi\big(\mathcal C\frac{\partial C^a}{\partial a_{bcd}}\big)_{\alpha\beta}-\frac23\big(\mathcal C\{\Gamma^{bcd}, C^a\}\big)_{\alpha\beta}+\frac{7}{10}\big(\mathcal C E^{a}{}_{e}\tilde\Gamma^{eb,cd}\big)_{\alpha\beta}+\frac{1}{60}\big(\mathcal C\Gamma^{bcd}\Gamma_* E^a\big)_{\beta\alpha}\right)p_{bcd}&\label{eq:linpsi}\\[1mm]
-\left(2\big(\mathcal C\Gamma^{bc} A\big)_{\alpha\beta}+4\big(\mathcal C\Gamma^{bc}\Gamma_* B\big)_{\alpha\beta}+8\big(\mathcal C\Gamma^b C^c\big)_{\alpha\beta}\right)p^{a}{}_{bc}+4\,e^{3\phi}p^{bc[d}\,\mathcal T^{a]e}\,a_{ebc}\big(\mathcal C\Gamma_d\Gamma_*\big)_{\alpha\beta}&\nonumber\\[1mm]
+2\,e^{2\phi}p_{bcd}\,\mathcal T_{ef}\,a^{aeb}\,a^{cdf}\big(\mathcal C\Gamma_*\big)_{\alpha\beta}+\frac13 p^{bcd}\,b^{aefg}\,(\mathcal C \Gamma_{bcdefg}\Gamma_*)_{\alpha\beta}-\frac14 p^{abc}\,b^{defg}(\mathcal C\Gamma_{bcdefg}\Gamma_*)_{\alpha\beta}&=0\nonumber\,,%\;\;\;\;\chi_{a}^{\alpha}\nonumber
 \end{flalign}
where $\tilde\Gamma^{eb,cd}$ and $b^{abcd}$ are respectively defined in \eqref{eq:Gammacombi} and \eqref{eq:babcde}. In the first two equations, antisymmetrization over $[abc]$ is understood. In the last one, which is implicitly contracted with $\chi_a^\alpha$, we have kept the currents explicit. The source terms, which do not involve Yukawa tensors, come from the variation of the gauge field in the $\sim p_{abc}p^{abc}$ kinetic term, but also from the variation of the topological term. In the latter we have dualized the gamma matrices using \eqref{eq:gammadual}.

Subsequently imposing supersymmetry of the Lagrangian \eqref{eq:LSUGRA} at order $\mathrm{g}^2$ leads to an additional set of quadratic relations among the Yukawa tensors, which we list below for completeness,
\begin{align}
40(\mathcal C B)_{\gamma\alpha}\,( B)^\gamma{}_\beta+4\,(\mathcal CC^a)_{\gamma\alpha}\,( C^a)^\gamma{}_\beta+\mathcal C_{\alpha\beta} \,V_{\text{pot}}&\nonumber\\[1mm]
-\frac{\mathrm i^{(s-1)} }{384}\,e^{4\phi}\,\varepsilon^{a_1\ldots a_{10}}\,\mathcal T^{bc}\,\mathcal T^{de}\,a_{ba_1a_2}\,a_{ca_3a_4}\,a_{a_5a_6a_7}\,a_{da_8a_9}\,(\mathcal C\Gamma_{a_{10}e}\Gamma_*)_{\alpha\beta}&=0\,,\\[4mm]
-40(\mathcal CBA)_{\alpha\beta}+2(\mathcal CE)_{[\alpha\gamma]}\,(B)^\gamma{}_\beta+ (\mathcal C E_a C^a)_{\alpha\beta}&\nonumber\\[1mm]
+\frac{\partial V_{\text{pot}}}{\partial \phi}\,(\mathcal C\Gamma_*)_{\alpha\beta}-e^{\phi}\frac{\partial V_{\text{pot}}}{\partial a^{abc}}\,(\mathcal C\Gamma^{abc}\Gamma_*)_{\alpha\beta}&\nonumber\\
+\frac{\mathrm i^{(s-1)} }{96}\,e^{4\phi}\,\varepsilon^{a_1\ldots a_{10}}\,\mathcal T^{bc}\,\mathcal T^{de}\,a_{ba_1a_2}\,a_{ca_3a_4}\,a_{a_5a_6a_7}\,a_{da_8a_9}\,(\mathcal C\Gamma_{a_{10}e})_{\alpha\beta}&=0\,,\\[4mm]
-4 (\mathcal C C^a A)_{\alpha\beta}- (\mathcal C B)_{\gamma\beta}\,( E^a)^\gamma{}_\alpha+2 (\mathcal C  E^{ab})_{(\alpha\gamma)}\,(C_b)^\gamma{}_\beta&\nonumber\\[1mm]
+\frac{\partial V_{\text{pot}}}{\partial \mathcal T^{bc}}\,\Big(\delta_{\epsilon}\mathcal{T}_{bc}\Big)^a_{\alpha\beta}-3\,e^\phi\frac{\partial V_{\text{pot}}}{\partial a_{abc}}\,( C\Gamma_{bc})_{\alpha\beta}&\nonumber\\[1mm]
+\frac{\mathrm i^{(s-1)}}{192}\,e^{4\phi}\,\varepsilon^{a_1\ldots a_{10}}\,\mathcal T^{bc}\,\mathcal T_{d}{}^{e}\,a_{ba_1a_2}\,a_{ca_3a_4}\,a_{a_5a_6a_7}\,a_{da_8a_9}\,(\mathcal C\Gamma_{[a_{10}}\Gamma_*)_{\alpha\beta}\,\delta_{e]}{}^{a}&=0\,,
\end{align}
where the scalar potential $V_{\text{pot}}$ was defined in \eqref{eq:Lpot}. The terms proportional to the Levi–Civita tensor arise from the variations of the gauge field hidden in the current in the topological term $\mathcal L_{\text{top}}$. After a lengthy algebra, one can verify that these relations are indeed satisfied.

\end{appendix}

\providecommand{\href}[2]{#2}\begingroup\raggedright\endgroup

%\bibliographystyle{utphys}
%\bibliography{refs}

\end{document}